\documentclass[prd,aps,eqsecnum,floatfix,nofootinbib,preprint,tightenlines]{revtex4-2}

\usepackage{latexsym}
\usepackage{graphicx}
\usepackage{multirow}
\usepackage{amsmath}
\usepackage[dvipsnames]{xcolor}

\def\bibi{\bibitem}

\let\inodot=\i

\def\a{\alpha}

\def\c{\chi}
\def\d{\delta}
\def\f{\phi}                    %       \varphi
\def\g{\gamma}
\def\h{\eta}
\def\i{\iota}

\def\m{\mu}
\def\n{\nu}
\def\o{\omega}
\def\p{\pi}                     % Also, \varpi
\def\r{\rho}                    %       \varrho
\def\s{\sigma}                  %       \varsigma
\def\t{\tau}

\def\D{\Delta}

\def\co{{\cal O}}

\def\cbo{{\,\raise-.15ex\Sc [\,}}                       % curly "
\def\ddt#1{{\buildrel {\hbox{\LARGE .\kern-2pt.}} \over {#1}}}% double dot-over
\def\ie{\mbox{\it i.e.}}
\def\eg{\mbox{\it e.g.}}
\def\etc{\mbox{\it etc.}}

\def\floatcaption#1#2{ \caption{ #2 \ [#1] \label{#1}} }
\def\floatcaption#1#2{ \caption{#2 \label{#1}} }

\def\bibi{\bibitem}    %% uncomment to suppresses citation labels

\def\ttl#1{{\it #1}}
\def\seef{{\it cf.\  }}

\mathchardef\mhyphen="2D

\begin{document}

\begin{boldmath}
\begin{center}
{\large{\bf
Data-driven estimates for light-quark-connected and strange-plus-disconnected
hadronic $g-2$ window quantities}
}\\[8mm]
Genessa Benton,$^a$
Diogo Boito,$^b$ Maarten Golterman,$^a$ Alex Keshavarzi,$^c$
Kim Maltman,$^{d,e}$
 and
Santiago Peris$^{a,f}$\\[8 mm]
$^a$Department of Physics and Astronomy, San Francisco State University,\\
San Francisco, CA 94132, USA
\\[5mm]
$^b$Instituto de F{\'\inodot}sica de S{\~a}o Carlos, Universidade de S{\~a}o
Paulo\\ CP 369, 13570-970, S{\~a}o Carlos, SP, Brazil
\\
[5mm]
$^c$Department of Physics and Astronomy, The University of Manchester,\\
Manchester M13 9PL, United Kingdom\\
[5mm]
$^d$Department of Mathematics and Statistics,
York University,\\  Toronto, ON Canada M3J~1P3
\\[5mm]
$^e$CSSM, University of Adelaide, Adelaide, SA~5005 Australia
\\[5mm]
$^f$Department of Physics and IFAE-BIST, Universitat Aut\`onoma de Barcelona,\\
E-08193 Bellaterra, Barcelona, Spain
\\[10mm]
\end{center}
\end{boldmath}

\begin{quotation}
A number of discrepancies have emerged between lattice computations
and data-driven dispersive evaluations of the RBC/UKQCD
intermediate-window-hadronic contribution to the muon anomalous
magnetic moment. It is therefore interesting to obtain data-driven
estimates for the light-quark-connected and strange-plus-disconnected
components of this window quantity, allowing for a more detailed
comparison between the lattice and data-driven approaches. The aim
of this paper is to provide these estimates, extending the analysis to several
other window quantities, including two windows designed
to focus on the region in which the two-pion contribution is dominant.
Clear discrepancies are observed for all light-quark-connected contributions considered, while good agreement with lattice results is found for strange-plus-disconnected contributions to the quantities for which corresponding lattice results exist.
The largest of these discrepancies is that for the 
RBC/UKQCD intermediate window, where, as previously reported, 
our data-driven result, $a_\m^{W1,{\rm lqc}}=198.9(1.1)\times 10^{-10}$,
is in significant tension with the results of~8 different 
recent lattice determinations.
Our strategy is the same as recently employed in obtaining data-driven
estimates for the light-quark-connected and strange-plus-disconnected
components of the full leading-order hadronic vacuum polarization
contribution to the muon anomalous magnetic moment.
Updated versions of those earlier results are also presented, for
completeness.
\end{quotation}

%%%%%%%%%%%%%%%%%%%%%%%%%%%
%\newpage
\section{\label{intro} Introduction}
%%%%%%%%%%%%%%%%%%%%%%%%%%%
There have been many developments since the publication of the first of the Fermilab
E989 measurements of the muon anomalous magnetic moment \cite{FNL,FNL2}
and the publication of the White-Paper (WP) Standard-Model (SM) estimate
\cite{review} that preceded it. A lattice computation of the
leading-order hadronic vacuum polarization (HVP) contribution
$a_\m^{\rm HVP}$ by the BMW collaboration resulted in a value
that would bring the total SM expectation much closer to the
experimental value \cite{BMW}, if one assumes that the $5\s$
discrepancy between the world average of the experimental
value\footnote{Essentially that of the Brookhaven E821 \cite{BNL}
and Fermilab E989 \cite{FNL,FNL2} experiments.}
and the WP value is due to the leading-order HVP contribution. In
addition, there now exist a large number of lattice
computations \cite{BMW,Aubin:2019usy,Lehner:2020crt,Wang:2022lkq,Aubin:2022hgm,Ce:2022kxy,ExtendedTwistedMass:2022jpw,Bazavov:2023has,Blum:2023qou} of
the RBC/UKQCD intermediate window quantity \cite{RBC} that are in
relatively good agreement with each other, but not in agreement with
data-driven estimates \cite{Coletal}.

The lattice computation of $a_\m^{\rm HVP}$ is carried out by breaking
down the total contribution into several building blocks. The primary
building blocks are the isospin-symmetric light- and strange-quark
connected and three-flavor disconnected parts, together with
smaller charm and bottom contributions (both of which also have connected and
disconnected components). Electromagnetic (EM) and strong-isospin-breaking
(SIB) effects (collectively, isospin-breaking (IB) effects) are
taken into account perturbatively, with corrections linear in the
fine-structure constant $\a$ and the up-down quark mass difference
$m_u-m_d$ sufficiently precise for the current desired level of accuracy.
In contrast, the data-driven dispersive approach is based on an analysis
of hadronic electro-production data on a channel-by-channel basis
($\p^+\p^-$, $\p^+\p^-\p^0$, \etc) up to squared hadronic invariant
masses, $s$, just below $4$ GeV$^2$ and inclusive data above that point.
To gain a better understanding of the emerging discrepancies, it is
useful to find out in more detail which of the lattice components are
most strongly in tension with data-driven estimates. For this, it is
necessary to reorganize the data-driven approach such as to provide
direct estimates for the building blocks that constitute the lattice-based
computation of $a_\m^{\rm HVP}$. It is, in addition, of interest to consider
auxillary lattice quantities that focus on the region in which the
two-pion contribution dominates, in order to further sharpen our
understanding of the source(s) of the lattice-dispersive discrepancy.

In recent work, we have obtained such data-driven estimates for the
isospin-symmetric light-quark connected (``lqc'' for short) \cite{lqc}
and strange connected plus (three-flavor-) disconnected contributions
\cite{spd} to  $a_\m^{\rm HVP}$. We refer to the latter as the strange
plus disconnected contribution in what follows, or ``s+lqd'' for short.
While the strange plus disconnected contribution was found to be in good
agreement with most lattice computations (with comparable errors), the
light-quark connected contribution differed significantly from the lattice
result of Ref.~\cite{BMW}.\footnote{At present, no other lattice computations
claim sufficient control over all systematic errors to make a meaningful
comparison.} This state of affairs provides a strong motivation for
obtaining similar light-quark connected and strange plus disconnected
data-driven estimates for various window quantities such as the RBC/UKQCD
intermediate window, where, as mentioned above, a growing number of
lattice collaborations finds values not in agreement with data-driven
estimates.

The aim of this paper is to provide data-driven values for a number of
light-quark connected and strange plus disconnected window quantities,
specifically, the original RBC/UKQCD intermediate window quantity, the
longer-distance window quantity proposed in Ref.~\cite{Aubin:2022hgm}, and
two window quantities introduced in Ref.~\cite{sumrule}, and compare
these with corresponding lattice values where available. The result for the
data-based light-quark connected contribution to the intermediate window
obtained with our method has already been reported in a recent
letter~\cite{Benton:2023dci}. Here we extend the results to other
windows, to the strange plus disconnected contributions and provide
several additional details.

The analysis of Refs.~\cite{lqc,spd} was based on the tabulated exclusive-mode
hadronic contributions to $a_\m^{\rm HVP}$ made available in
Refs.~\cite{Davier:2019can} (DHMZ) and \cite{KNT19} (KNT). Such a tabulation
is not publicly available for any of the window quantities considered
in this paper. One therefore needs access to the exclusive-mode spectral
distributions for all hadronic channels contributing to the HVP
in order to evaluate the channel-by-channel
contributions to these window quantities, which then, following the
strategy of Refs.~\cite{lqc,spd}, can be used to obtain the light-quark
connected and strange plus disconnected contributions to these window
quantities. In this paper, we will use the
exclusive-mode spectral distributions and covariances from the 2019 KNT
analysis (KNT19 in what follows). As analogous exclusive-mode
information from DHMZ is not available to us, we are unable to present
estimates based on DHMZ data. In this respect the analysis presented
here differs from that in Refs.~\cite{lqc,spd}, in which both KNT19-
and DHMZ-based values for the light-quark connected and strange plus
disconnected contributions to $a_\m^{\rm HVP}$ were obtained.

Of course, both the KNT and DHMZ data contain EM and SIB effects.
These have to be estimated and subtracted in order to arrive at purely
hadronic and isospin-symmetric estimates for the various window
quantities. Both types of corrections have been recently considered in
Refs.~\cite{Colangelo:2022prz,Hoferichter:2022iqe,Hoferichter:2023sli,Hoferichter:2023bjm}. 
The existence of as-yet-unquantified
exclusive-mode contributions (see, {\it e.g.}, the discussions in Refs.~\cite{Hoferichter:2022iqe,Hoferichter:2023sli}
and the appendix of Ref.~\cite{lqc}) means insufficient experimental information is currently
available to extract inclusive EM corrections, and we thus choose to rely
on available lattice estimates for the EM corrections we employ;
these will be the only lattice data we will make use of in this paper.
As we will see, these corrections are very small. Additional IB
corrections, in which a combination of EM and SIB effects occur, are
amenable to data-driven treatment. Useful information on the sum of
EM and SIB contributions in the two-pion channel, which is expected to
be dominated by its SIB component, is provided by Ref.~\cite{Colangelo:2022prz}.
We will use these results in assessing and subtracting this sum.
Other such EM+SIB corrections will also be discussed below.
We will use a definition of QCD in the isospin limit in which
the pion mass is equal to the physical $\p^0$ mass. To first order
in IB, \ie, to $\co(\a)$ and $\co(m_u-m_d)$, this is
sufficient to define an unambiguous split between EM and SIB corrections.

This paper is organized as follows.
In Sec.~\ref{review} we define the window quantities of interest in this
paper (Sec.~\ref{windows}), and remind the reader how to relate the lqc and
s+lqd parts of the EM spectral function to its isospin components. In
Sec.~\ref{implement} we present our results, based on the KNT19 data, for the
exclusive-channel contributions to the lqc and s+lqd parts of our window
quantities in the region below $s=(1.937)^2$~GeV$^2$, treating first the modes
for which the isospin can be identified through $G$-parity (Sec.~\ref{Gpar}), and
then the isospin-ambiguous modes (Sec.~\ref{amb}). Above $s=(1.937)^2$~GeV$^2$
we employ perturbation theory, which is discussed in Sec.~\ref{pertth}.
IB corrections are discussed in detail in Sec.~\ref{EMSIB}. We then compare our
results with recent lattice results for the same quantities in
Sec.~\ref{comp}, and conclude in Sec.~\ref{concl}.
There are two appendices, one tabulating intermediate results in addition to
those explained in the main text, and one with a more detailed discussion of the
$\p^0\g$ and $\eta\g$ exclusive modes.

\section{\label{review} Review}
We define the window quantities we will consider in this paper in
Sec.~\ref{windows}, and recall how to write the light-quark connected
and strange plus disconnected spectral functions in terms of the
$I=1$ and $I=0$ components.

\subsection{\label{windows} Windows}
The leading-order HVP contribution $a_\m^{\rm HVP}$ to $a_\m$ can be written,
in dispersive form, as \cite{Brodsky:1967sr,Lautrup:1968tdb,Gourdin:1969dm}
\begin{equation}
\label{amuHVP}
a_\m^{\rm HVP}=\frac{4\alpha^2m_\mu^2}{3}
\int_{m_\pi^2}^\infty ds\, \frac{\hat{K}(s)}{s^2} \r_{\rm EM}(s)\ ,
\end{equation}
where $m_\p$ is the neutral pion mass. Here $\r_{\rm EM}(s)$ is the
inclusive EM-current hadronic spectral function
\begin{eqnarray}
\label{rdefn}
\r_{\rm EM}(s)&=&\frac{1}{12\p^2}R(s)\ ,\\
R(s)&=&\frac{3s}{4\pi \alpha^2}\,
\sigma^{(0)} [e^+ e^-\rightarrow \mbox{hadrons}(+\gamma)]\ ,\nonumber
\end{eqnarray}
where $R(s)$ is the $R$ ratio obtained from the bare inclusive hadronic
electroproduction cross section
$\s^{(0)}[e^+ e^-\rightarrow \mbox{hadrons}(+\g)]$, and  $\hat{K}(s)$
is a known smoothly varying kernel with $\hat{K}(4m_\p^2)\approx 0.63$
at the two-pion threshold and $\lim_{s\to\infty}\hat{K}(s)=1$.\footnote{For
various explicit expressions, see \eg\ Ref.~\cite{review}.} Equivalently,
$a_\m^{\rm HVP}$ can be expressed in terms of the Euclidean-time
two-point correlation function \cite{Bernecker:2011gh}
\begin{equation}
\label{defC}
C(t)=\frac{1}{3}\sum_{i=1}^3\int d^3x \langle j_i^{\rm EM}(\vec{x},t)
j_i^{\rm EM}(0)\rangle=\frac{1}{2}\int_{m_\p^2}^\infty ds\,\sqrt{s}\,
e^{-\sqrt{s}t}\,\r_{\rm EM}(s)\quad (t>0)\ ,
\end{equation}
of the EM current, $j_i^{\rm EM}(\vec{x},t)$, as
\begin{equation}
\label{amuHVPtimemom}
a_\m^{\rm HVP}=2\int_0^\infty dt\,w(t) C(t)\ ,
\end{equation}
where $w(t)$ is a known function related to $\hat{K}(s)$ by\footnote{For
an explicit expression as well as an approximation accurate to better than
one part in $3\times 10^{-6}$, see Ref.~\cite{DellaMorte:2017dyu}.}
\begin{equation}
\label{wvsK}
\frac{\hat{K}(s)}{s^2}=\frac{3\sqrt{s}}{4\a^2 m_\m^2}
\int_0^\infty dt\,w(t)\,e^{-\sqrt{s}t}\ .
\end{equation}
We note that $C(t)$ has a $\d$-function singularity at $t=0$, but this
does not contribute to Eq.~(\ref{amuHVPtimemom}) since $w(t)\sim t^4$ for $t\to 0$.

Our first class of window quantities is defined by inserting the window
function~\cite{RBC}
\begin{equation}
\label{window}
W(t;t_0,t_1;\D)=\frac{1}{2}\left(\tanh\frac{t-t_0}{\D}-\tanh\frac{t-t_1}{\D}\right)
\end{equation}
with $t_1>t_0>0$ into Eq.~(\ref{amuHVPtimemom}):
\begin{eqnarray}
\label{amuW}
a_\m^W(t_0,t_1;\D)&=&2\int_0^\infty dt\,W(t;t_0,t_1;\D)\,w(t) C(t)\\
&=&\int_{m_\p^2}^\infty ds\,\r_{\rm EM}(s)\,\sqrt{s}
\int_0^\infty dt\,W(t;t_0,t_1;\D)\,w(t)\,e^{-\sqrt{s}t}
\nonumber\\
&=& \frac{4\alpha^2m_\mu^2}{3}\int_{m_\p^2}^\infty ds\,
\frac{\hat{K}(s)}{s^2} \,\widetilde{W}(s;t_0,t_1;\D)\,\r_{\rm EM}(s)\ ,\nonumber
\end{eqnarray}
where
\begin{equation}
\label{windowint}
\widetilde{W}(s;t_0,t_1;\D)=\frac{\int_0^\infty dt\,W(t;t_0,t_1;\D)\,
w(t)\,e^{-\sqrt{s}t}}{\int_0^\infty dt\,w(t)\,e^{-\sqrt{s}t}}
\end{equation}
is the window function in $s$-space.

In what follows, we will consider two window quantities of this type,
$a_\mu^{W1}$ and $a_\mu^{W2}$, corresponding to the window functions
$W1$ and $W2$ obtained using the choices
\begin{subequations}
\label{Ws}
\begin{eqnarray}
&W1:\qquad &t_0=0.4~\mbox{fm}\ ,\quad t_1=1.0~\mbox{fm}\ ,
\quad\D=0.15~\mbox{fm}\ ,\label{Wsa}\\
&W2:\qquad &t_0=1.5~\mbox{fm}\ ,\quad t_1=1.9~\mbox{fm}\ ,
\quad\D=0.15~\mbox{fm}\ ,\label{Wsb}
\end{eqnarray}
\end{subequations}
for the external parameters, $t_0$, $t_1$ and $\Delta$, in Eq.~(\ref{window}).
The first of these, $a_\mu^{W1}$, is the RBC/UKQCD intermediate window
quantity introduced in Ref.~\cite{RBC}. The second, $a_\m^{W2}$, is the
alternate longer distance ``intermediate'' window quantity introduced
in Ref.~\cite{Aubin:2022hgm} designed to be better suited to treatment using
chiral perturbation theory.

We will also consider window quantities $I_{\widehat{W}}$,
\begin{equation}
\label{windowsr}
I_{\widehat{W}}=\int_{m_\p^2}^\infty ds\,\widehat{W}(s)\,\r_{\rm EM}(s)\ ,
\end{equation}
involving weights, $\widehat{W}(s)$, defined directly as functions of
$s$ and having the form\footnote{The design of such weights
in Ref.~\cite{sumrule} was inspired by the work of Ref.~\cite{Hansen:2019idp}.}
\begin{equation}
\label{What}
\widehat{W}(s;\{t_i\},\{x_i\})=\frac{1}{2}\sum_{i=1}^n x_i \sqrt{s}
e^{-t_i\sqrt{s}}\ ,
\end{equation}
with $\{ t_i\}$ a set of positive numbers. Using Eq.~(\ref{defC}) it follows that
\begin{equation}
\label{sumrule}
I_{\widehat{W}}=\sum_{i=1}^n x_i C(t_i)\ ,
\end{equation}
thus providing a sum rule that allows for a comparison between the spectral
integral $I_{\widehat{W}}$ and the weighted sum of $n$ values of the
correlation function $C(t)$ that can be computed on the lattice
\cite{sumrule}. By choosing $n$ and the sets $\{t_i\}$ and $\{x_i\}$
judiciously, weights can be constructed to focus on particular regions
of interest in $s$. A key advantage of the form Eq.~(\ref{What}) is that it
turns out to be possible to obtain such localization in $s$ using sets
$\{ t_i\}$ which entirely avoid the large-$t$ region in which lattice
$C(t)$ errors deteriorate \cite{sumrule}, thus simultaneously controlling
errors on the lattice side of the sum rule Eq.~(\ref{sumrule}). The lattice
errors corresponding to four such choices, $W^\prime_{15}$, $W^\prime_{25}$,
$\widehat{W}_{15}$ and $\widehat{W}_{25}$, all focusing on the region
around the $\r$ peak, were investigated in Ref.~\cite{sumrule}, using
available light-quark connected $C(t)$ results, with $\widehat{W}_{15}$
and $\widehat{W}_{25}$ found to produce the smallest relative lattice
errors. We thus focus on these cases in what follows. Both involve an
$n=5$-fold sum, and the choice
\begin{equation}
\label{ts}
\{t_i\}=\{3,\ 6,\ 9,\ 12,\ 15\}\ \mbox{GeV}^{-1}
\end{equation}
for the set $\{ t_i\}$.
The weight $\widehat{W}_{15}$ is then obtained by choosing
\begin{equation}
\label{W15}
\{x_i^{(15)}\}= \{-78.8487,\ 5688.30,\ 2223.96,\ -36638.0,\ 8047.38\}\ ,
\end{equation}
and $\widehat{W}_{25}$ by choosing
\begin{equation}
\label{W25}
\{x_i^{(25)}\}= \{44.8916,\ 590.933,\ -3373.53,\ 3716.86,\ 8047.38,
\ 879.149\}\ .
\end{equation}
With the inclusive spectral function $\r_{\rm EM}(s)$ one can obtain
data-driven values for $a_\m^{W1}$, $a_\m^{W2}$, $I_{\widehat{W}_{15}}$
and $I_{\widehat{W}_{25}}$. For $a_\m^{W1}$ (and a number of other
weights which we will not consider here) this was done in Ref.~\cite{Coletal}.
Here we are interested in obtaining, for each of the weights we consider,
the light-quark connected and strange plus disconnected building blocks.

\subsection{\label{lqcspd} Light-quark connected and strange plus disconnected spectral functions}
First, we review the ingredients of the basic idea from Refs.~\cite{lqc,spd}.
The decomposition of the three-flavor EM current into its isospin
$I=1$ and $I=0$ parts, produces related decompositions of $C(t)$ and
$\r_{\rm EM}(s)$, into pure $I=1$, pure $I=0$ and mixed-isospin parts
\begin{eqnarray}
\label{Idecomp}
C(t)&=&C^{I=1}(t)+C^{\rm MI}(t)+C^{I=0}(t)\ ,\\
\r_{\rm EM}(s)&=&\r_{\rm EM}^{I=1}(s)+\r_{\rm EM}^{\rm MI}(s)
+\r_{\rm EM}^{I=0}(s)\ ,\nonumber
\end{eqnarray}
where in an isospin symmetric world the mixed-isospin (MI) components
vanish. Weighted integrals over $\r_{\rm EM}$, of course, inherit this
same decomposition.

In the isospin limit, the $I=0$ contribution to the light-quark
connected (lqc) part of $C(t)$ is exactly $1/9$ times the corresponding
$I=1$ contribution. The strange (connected plus disconnected)
plus light-quark disconnected (s+lqd) contribution
is, similarly, the difference of the $I=0$ contribution and $1/9$ times
the $I=1$ contribution. The lqc and s+lqd window quantities considered
in this paper, which are of the form~(\ref{amuW}) with $W=W1$ or
$W=W2$, or of the form~(\ref{windowsr}) with $\widehat{W}=\widehat{W}_{15}$
or $\widehat{W}=\widehat{W}_{25}$,
are thus given, in the isospin limit, by the expressions
\begin{eqnarray}
\label{amuWsplit}
a_\m^{W,\rm lqc}&=&\frac{4\alpha^2m_\mu^2}{3}\int_{m_\p^2}^\infty ds\,
\frac{\hat{K}(s)}{s^2} \,\widetilde{W}(s;t_0,t_1;\D)\,
\r_{\rm EM}^{\rm lqc}(s)\ ,\\
a_\m^{W,\rm s+lqd}&=&\frac{4\alpha^2m_\mu^2}{3}\int_{m_\p^2}^\infty ds\,
\frac{\hat{K}(s)}{s^2} \,\widetilde{W}(s;t_0,t_1;\D)\,
\r_{\rm EM}^{\rm s+lqd}(s)\ ,\nonumber
\end{eqnarray}
and
\begin{eqnarray}
\label{Isplit}
{\cal{I}}_{\widehat{W}}^{\rm lqc}&=&\int_{m_\pi^2}^\infty ds \,
\widehat{W}(s)\, \rho^{\rm lqc}_{\rm EM}(s)\ ,\\
{\cal{I}}_{\widehat{W}}^{\rm s+lqd}&=&\int_{m_\pi^2}^\infty ds \,
\widehat{W}(s)\, \rho^{\rm s+lqd}_{\rm EM}(s)\ ,\nonumber
\end{eqnarray}
where
\begin{equation}
\label{lqcspec}
\r^{\rm lqc}_{\rm EM}(s)=\frac{10}{9}\r_{\rm EM}^{I=1}(s)
\end{equation}
and
\begin{equation}
\label{spdspec}
\r^{\rm s+lqd}_{\rm EM}(s)=\r_{\rm EM}^{I=0}(s)-\frac{1}{9}
\r_{\rm EM}^{I=1}(s)=\r_{\rm EM}(s)-\frac{10}{9}\r_{\rm EM}^{I=1}(s)\ .
\end{equation}
The evaluation of the lqc and s+lqd parts of the four window quantities
defined in Sec.~\ref{windows} thus requires an identification of the separate
$I=1$ and $I=0$ components of $\rho_{\rm EM}(s)$. The separation of
contributions from all hadronic exclusive modes can, assuming isospin
symmetry, be accomplished, up to $\sqrt{s}=1.937$ GeV, using KNT19
exclusive-mode data, as described in the following section. We will then,
in Sec.~\ref{EMSIB}, discuss the EM and SIB corrections to our values for
the lqc and s+lqd parts of our window quantities.

\section{\label{implement} Implementation}
In this section, we obtain the lqc and s+lqd parts for all four window
quantities, postponing until Sec.~\ref{EMSIB} a consideration of EM and SIB
corrections. In Sec.~\ref{Gpar} we collect exclusive-mode contributions from
modes which are $G$-parity eigenstates and hence have an unambiguous $I=0$
or $1$ assignment. The separation of contributions from modes which are
not $G$-parity eigenstates into separate $I=0$ and $I=1$ components is
detailed in Sec.~\ref{amb}.

%\newpage
\subsection{\label{Gpar} Modes with unambiguous isospin}
As in Ref.~\cite{spd}, we take advantage of the fact that exclusive modes
with positive (negative) $G$-parity have $I=1$ ($I=0$). The contributions
of such modes to the $I=1$ and $I=0$ parts of $a_\m^{W1}$ are shown in
Table~\ref{tab1}. Analogous tables are provided for the other window
quantities in App.~\ref{otherweights}. From these results, we obtain the
following $G$-parity unambiguous contributions to our window quantities:
\begin{eqnarray}
\label{Gparlqc}
\left[a_\m^{W1,\rm lqc}\right]_{G\rm{\mhyphen par}} &=&\frac{10}{9}
\times 168.24(72)\times 10^{-10}=186.94(80)\times 10^{-10}\ ,\nonumber\\
\left[a_\m^{W2,\rm lqc}\right]_{G\rm{\mhyphen par}} & =&\frac{10}{9}
\times 85.05(33)\times 10^{-10}=94.50(36)\times 10^{-10}\ ,\nonumber\\
\left[I_{\widehat{W}_{15}}^{\rm lqc}\right]_{G\rm{\mhyphen par}} &=&
\frac{10}{9}\times 39.37(14)\times 10^{-2}=43.75(15)\times 10^{-2}\
,\nonumber\\
\left[I_{\widehat{W}_{25}}^{\rm lqc}\right]_{G\rm{\mhyphen par}} &= &
\frac{10}{9}\times 67.29(30)\times 10^{-3}=74.76(34)\times 10^{-3}\ ,
\end{eqnarray}
and 
\begin{eqnarray}
\label{Gparspd}
\left[a_\m^{W1,\rm s+lqd}\right]_{G\rm{\mhyphen par}} &= &\left[
20.69(37)-\frac{168.24(72)}{9}\right]\times 10^{-10}
=1.99(38)\times 10^{-10}\ ,
\nonumber\\
\left[a_\m^{W2,\rm s+lqd}\right]_{G\rm{\mhyphen par}} &= &\left[
6.23(14)-\frac{85.05(33)}{9}\right]\times 10^{-10}
=-3.22(14)\times 10^{-10}\ , \nonumber\\
\left[I_{\widehat{W}_{15}}^{\rm s+lqd}\right]_{G\rm{\mhyphen par}} &=&
\left[3.77(08)-\frac{39.37(14)}{9}\right]\times 10^{-2}
=-0.609(84)\times 10^{-2}\ ,\nonumber\\
\left[I_{\widehat{W}_{25}}^{\rm s+lqd}\right]_{G\rm{\mhyphen par}} &= &
\left[8.92(16)-\frac{67.29(30)}{9}\right]\times 10^{-3}
=1.45(16)\times 10^{-3}\ ,
\end{eqnarray}
\begin{table}[t]
\begin{center}
\begin{tabular}{lr|lr}
\hline
$I=1$ modes $X$&$[a_\mu^{W1}]_X\times 10^{10}$&$I=0$ modes $X$&
$[a_\mu^{W1}]_X\times 10^{10}$\\
\hline
low-$s$ $\pi^+ \pi^-$& 0.02(00)\quad& low-$s$ $3\pi$& 0.00(00)\quad\\
$\pi^+ \pi^-$& 144.13(49)\quad& $3\pi$& 18.69(35)\quad\\
$2\pi^+ 2\pi^-$& 9.29(13)\quad&  $2\pi^+2\pi^-\pi^0$ (no $\omega$,
$\eta$)& 0.61(06)\quad\\
$\pi^+ \pi^- 2\pi^0$& 11.94(48)\quad& $\pi^+ \pi^- 3\pi^0$ (no $\eta$)&
0.39(07)\quad\\
$3\pi^+ 3\pi^-$ (no $\omega$)& 0.14(01)\quad& $3\pi^+3\pi^- \pi^0$ (no
$\omega$, $\eta$)& 0.00(00)\quad\\
$2\pi^+2\pi^-2\pi^0$ (no $\eta$)& 0.83(11)\quad& $\eta \pi^+\pi^-\pi^0$
(no $\omega$)&0.44(05)\quad\\
$\pi^+\pi^- 4\pi^0$ (no $\eta$)& 0.13(13)\quad& $\eta \omega$
& 0.19(01)\quad\\
$\eta \pi^+ \pi^-$& 0.85(03)\quad& $\omega (\rightarrow npp )2\pi$&
0.08(01)\quad\\
$\eta 2\pi^+ 2\pi^-$& 0.05(01)\quad& $\omega 2\pi^+ 2\pi^-$&
0.00(00)\quad\\
$\eta \pi^+\pi^- 2\pi^0$& 0.07(01)\quad& $\eta \phi$& 0.25(01)\quad\\
$\omega (\rightarrow \pi^0\gamma)\pi^0$& 0.53(01)\quad& $\phi \rightarrow
unaccounted$ &  0.02(02)\quad\\
$\omega (\rightarrow npp)3\pi$& 0.10(02)\quad& \\
$\omega \eta \pi^0$& 0.15(03)\quad&\\
\hline
TOTAL:& 168.24(72)\quad&TOTAL:& 20.69(37)\quad\\
\hline
\end{tabular}
\floatcaption{tab1}{\it $G$-parity-unambiguous exclusive-mode
contributions to $a_\m^{W1}$ for $\sqrt{s}\leq 1.937$~{\rm GeV} using
KNT19 exclusive-mode data. Entries in units of $10^{-10}$. The
notation ``npp'' is KNT shorthand for ``non-purely-pionic.''}
\end{center}
\end{table}

\subsection{\label{amb} Modes with ambiguous isospin}
We now consider those exclusive modes which are not $G$-parity eigenstates,
and thus have no definite isospin. Associated contributions to the
various window quantities thus, in general, have both $I=1$ and $I=0$
components, which must be separated to complete determinations
of the corresponding lqc and s+lqd contributions. Modes of
this type in the KNT19 exclusive-mode region are
$\pi^0\gamma$, $\eta\gamma$, $N\bar{N}$ and those containing
a $K\bar{K}$ pair.

For the numerically most important modes of the latter type,
$K\bar{K}$ and $K\bar{K}\pi$, the $I=0/I=1$ separation is
facilitated by use of additional experimental input
($\tau^-\rightarrow K^-K^0\nu_\tau$ data \cite{BaBar:2018qry} for
$K\bar{K}$ and BaBar Dalitz plot analysis results \cite{BaBar:2007ceh}
for $K\bar{K}\pi$), following the strategy of Ref.~\cite{spd}, outlined
briefly below.

For the modes $P\gamma$ with $P=\pi^0,\, \eta$, the cross sections in
the KNT19 exclusive-mode region of relevance to the weighted integrals
considered in this paper are strongly dominated by contributions involving
intermediate $V=\rho$, $\omega$ and $\phi$ meson exchange.
Vector-meson-dominance (VMD) representations of the $e^+ e^-\rightarrow
P\gamma$ cross sections, which require as input only $m_P$, the vector
meson masses and widths and known experimental $V\rightarrow e^+ e^-$
and $V\rightarrow P\gamma$ decay widths, turn out to saturate the
corresponding KNT19 exclusive-mode integrals considered here, confirming
the reliability of the VMD representation. The $I=0$, $I=1$ and MI
contributions to those integrals can thus be determined from the parts
of the VMD representations involving the squared modulus of the sum of
$\omega$ and $\phi$ contributions to the amplitude, the squared modulus
of the $\rho$ contribution to the amplitude, and the interference terms
between those contributions, respectively. The VMD representation of
the cross sections, together with the numerical values of the required
external inputs, are detailed in Appendix~\ref{PgammaVMDdecomp}.
As we will see below, the $I=0$ contribution dominates these
integrals for all four window quantities considered in this paper.

For the (much smaller) contributions from other (``residual'')
$G$-parity-ambiguous exclusive-modes, $X$, where additional
external experimental input is not available, we follow
Ref.~\cite{spd} in employing a ``maximally conservative'' assessment
of the $I=1$/$I=0$ separation, based on the observation that the
$I=1$ part of the mode $X$ contribution to $\r_{\rm EM}(s)$
necessarily lies between $0$ and the full contribution. The mode
$X$ lqc and s$+$lqd contributions to $\r_{\rm EM}(s)$ then necessarily
lie in the ranges:
\begin{eqnarray}
\label{maxconservrhoem}
\left[\rho^{\rm lqc}_{\rm EM}\right]_{X}&=&\left(\frac{5}{9}\pm
\frac{5}{9}\right)\left[\rho_{\rm EM}\right]_{X}\ ,\\
\left[\rho^{\rm s+lqd}_{\rm EM}\right]_{X}&=&\left(\frac{4}{9}\pm
\frac{5}{9}\right)\left[\rho_{\rm EM}\right]_{X}\ .
\nonumber
\end{eqnarray}

These bounds produce related ``maximally conservative'' bounds
for weighted exclusive-mode lqc and s+lqd integrals involving
weights, $W$, having fixed sign in the region of nonzero
$\left[\rho_{\rm EM}(s)\right]_{X}$:{\footnote{The weights
$W1(s)$, $W2(s)$ and $\widehat{W}_{25}(s)$ are positive for all $s$.
$\widehat{W}_{15}(s)$, however, crosses zero, becoming small and
negative above $s\simeq 2.04$ GeV$^2$. We have checked that
non-negligible KNT19 results for $\left[\rho_{\rm EM}(s)\right]_{X}$
for all modes $X$ for which the maximally conservative assessment has
been employed lie in regions with fixed sign for $\widehat{W}_{15}(s)$,
making the bounds of Eq.~(\ref{maxconserv}) valid for $\widehat{W}_{15}$
as well.}}
\begin{eqnarray}
\label{maxconserv}
\left[a_\m^{W,\rm lqc}\right]_X&=&\left(\frac{5}{9}\pm \frac{5}{9}\right)
\left[a_\m^W\right]_X\ ,\\
\left[a_\m^{W,\rm s+lqd}\right]_X&=&\left(\frac{4}{9}\pm \frac{5}{9}\right)
\left[a_\m^W\right]_X\ .
\nonumber
\end{eqnarray}
The maximally conservative $I=1$/$I=0$ separation bounds, while
generally valid, would, if applied to all $G$-parity-ambiguous
modes, produce errors so large that our estimates for the lqc and
s+lqd parts of the window quantities would be uninteresting. Fortunately,
for those modes where this would be an issue, more experimental
information is available, allowing us to dramatically reduce the
uncertainty on the $I=1$/$I=0$ separation.

Let us first consider the $K\bar{K}$ modes, $K^+K^-$ and $K^0\bar{K}^0$.
Independent information on the $K\bar{K}$ contribution to the $I=1$
spectral function is available from data on the differential distribution
for the decay $\t\to K^-K^0\n_\t$ measured by BaBar \cite{BaBar:2018qry}.
Using the CVC (conserved vector current) relation, these results can be
converted into the $I=1$ $e^+e^-\to K\bar{K}$ contribution to
$R(s)$.\footnote{Isospin-breaking corrections to the CVC relation
are negligible for our purposes.} Using the results of
Ref.~\cite{BaBar:2018qry} up to $s=2.7556$~GeV$^2$, and a
KNT19-based maximally conservative assessment of $I=1$ contributions
for $s$ from $2.7556$~GeV$^2$ to $s=(1.937)^2$~GeV$^2$, we find,
following the steps of Sec.~IV.B of Ref.~\cite{spd}, the results
\begin{eqnarray}
\label{KKbarlqc}
\left[a_\m^{W1,\rm lqc}\right]_{K\bar{K}}&=&\frac{10}{9}\left(0.465(29)
+0.055(55)\right)\times 10^{-10} =0.578(69)\times 10^{-10}
\ ,\\
\left[a_\m^{W2,\rm lqc}\right]_{K\bar{K}}&=&\frac{10}{9}\left(0.0241(17)
+0.00035(35)\right)\times 10^{-10}=0.0271(20)\times 10^{-10}
\ ,\nonumber\\
\left[I_{\widehat{W}_{15}}^{\rm lqc}\right]_{K\bar{K}}&=&\frac{10}{9}
\left(0.0158(12)-0.0023(23)\right)\times 10^{-2}=0.0150(29)
\times 10^{-2}
\ ,\nonumber\\
\left[I_{\widehat{W}_{25}}^{\rm lqc}\right]_{K\bar{K}}&=&\frac{10}{9}
\left(0.209(13)+0.024(24)\right)\times 10^{-3}=0.259(30)\times 10^{-3}
\ ,\nonumber
\end{eqnarray}
where in each case the first number in parentheses is the contribution
from the BaBar $\tau$ data in the region $s\le 2.7556$~GeV$^2$, and the
second number is the result of the maximally conservative assessment
of the contribution from $s=2.7556$~GeV$^2$ to $(1.937)^2$ GeV$^2$,
obtained using KNT19 data.

For the s+lqd parts, using the second equation of Eq.~(\ref{spdspec}), we find
\begin{eqnarray}
\label{KKbarspd}
\left[a_\m^{W1,\rm s+lqd}\right]_{K\bar{K}}&=&\left(19.13(15)
-0.578(69)\right)\times 10^{-10}=18.55(17)\times 10^{-10}
\ ,\\
\left[a_\m^{W2,\rm s+lqd}\right]_{K\bar{K}}&=&\left(2.612(21)
-0.0271(20)\right)\times 10^{-10}=2.585(21)\times 10^{-10}
\ ,\nonumber\\
\left[I_{\widehat{W}_{15}}^{\rm s+lqd}\right]_{K\bar{K}}&=&
\left(2.213(18)-0.015(29)\right)\times 10^{-2}
=2.198(18)\times 10^{-2}
\ ,\nonumber\\
\left[I_{\widehat{W}_{25}}^{\rm s+lqd}\right]_{K\bar{K}}&=&
\left(8.642(70)-0.259(30)\right)\times 10^{-3}
=8.383(76)\times 10^{-3}
\ ,\nonumber
\end{eqnarray}
where the first number in parentheses in each case is the full
$K^+K^-$ plus $K^0\bar{K}^0$ contribution from the KNT19 data, and
the second number is that of Eq.~(\ref{KKbarlqc}). By comparing
Eq.~(\ref{KKbarlqc}), which is pure $I=1$, and Eq.~(\ref{KKbarspd}),
which is dominated by $I=0$, we see that clearly the $K\bar{K}$ channels
are dominated by $I=0$. This is expected because of the $I=0$ $\f$ resonance.

Our next case to consider is that of the $K\bar{K}\p$ modes. The separation
of the corresponding cross sections into their $I=0$ and $I=1$ parts was
carried out by BaBar in Ref.~\cite{BaBar:2007ceh}, and, as in Ref.~\cite{spd}, we
obtain the $I=1$ (and hence the lqc) contributions, up to
$s=(1.937)^2\ {\rm GeV}^2$,
\begin{eqnarray}
\label{KKbarpilqc}
\left[a_\m^{W1,\rm lqc}\right]_{K\bar{K}\p}&=&\frac{10}{9}\left[
a_\m^{W1,I=1}\right]_{K\bar{K}\p}=0.521(86)\times 10^{-10}
\ ,\\
\left[a_\m^{W2,\rm lqc}\right]_{K\bar{K}\p}&=&\frac{10}{9}\left[
a_\m^{W2,I=1}\right]_{K\bar{K}\p}=0.0060(10)\times 10^{-10}
\ ,\nonumber\\
\left[I_{\widehat{W}_{15}}^{\rm lqc}\right]_{K\bar{K}\p}&=&\frac{10}{9}
\left[I_{\widehat{W}_{15}}^{I=1}\right]_{K\bar{K}\p}=-0.0152(25)
\times 10^{-2}
\ ,\nonumber\\
\left[I_{\widehat{W}_{25}}^{\rm lqc}\right]_{K\bar{K}\p}&=&\frac{10}{9}
\left[I_{\widehat{W}_{25}}^{I=1}\right]_{K\bar{K}\p}=0.229(38)
\times 10^{-3}
\ .\nonumber
\end{eqnarray}
For the s+lqd $K\bar{K}\p$ contributions from the same region,
we find
\begin{eqnarray}
\label{KKbarpispd}
\left[a_\m^{W1,\rm s+lqd}\right]_{K\bar{K}\p}&=&\left(1.714(74)
-0.521(86)\right)\times 10^{-10}=1.19(11)\times 10^{-10}
\ ,\\
\left[a_\m^{W2,\rm s+lqd}\right]_{K\bar{K}\p}&=&\left(0.01846(82)-
0.0060(10)\right)\times 10^{-10}=0.0124(13)\times 10^{-10}
\ ,\nonumber\\
\left[I_{\widehat{W}_{15}}^{\rm s+lqd}\right]_{K\bar{K}\p}&=&\left(
-0.0536(24)+0.0152(25)\right)\times 10^{-2}=-0.0384(35)
\times 10^{-2}\ ,\nonumber\\
\left[I_{\widehat{W}_{25}}^{\rm s+lqd}\right]_{K\bar{K}\p}&=&\left(
0.753(33)-0.229(38)\right)\times 10^{-3}=0.524(50)
\times 10^{-3} \ ,\nonumber
\end{eqnarray}
where the first number in parentheses in each case is the full
$K\bar{K}\p$ contribution obtained using KNT19 data, and the second
number is that of Eq.~(\ref{KKbarpilqc}).

A small improvement, relative to the maximally conservative
assessment~(\ref{maxconserv}), can also be obtained for contributions
from the $K\bar{K}2\p$ modes by making use of the measured
$e^+e^-\to\f\p\p$ mode cross sections \cite{BaBar:2011btv}, which
allow the purely $I=0$ contribution resulting from
$e^+ e^- \to\phi [\rightarrow K\bar{K}]\pi\pi$ to be subtracted
from the KNT19 $K\bar{K}2\pi$ total and the maximally conservative
separation into $I=1$ and $I=0$ components applied only to the
remainder.  Following Ref.~\cite{spd}, we find\footnote{Note that the values,
 $49.2\%$ and $83.2\%$, for the $\f\to K^+K^-$ and two-mode
$\f\to K\bar{K}$ branching fractions, used in Ref.~\cite{spd} to update
external input employed in the original BaBar analysis, have been
further updated to current PDG \cite{Workman:2022ynf} values,
$49.1\%$ and $83.0\%$, respectively.}
\begin{eqnarray}
\label{KKbarpipilqc}
\left[a_\m^{W1,\rm lqc}\right]_{K\bar{K}2\p}&=&0.60
(60)\times 10^{-10}
\ ,\\
\left[a_\m^{W2,\rm lqc}\right]_{K\bar{K}2\p}&=&0.0031
(31)\times 10^{-10}
\ ,\nonumber\\
\left[I_{\widehat{W}_{15}}^{\rm lqc}\right]_{K\bar{K}2\p}&=&-0.028
(28)\times 10^{-2}
\ ,\nonumber\\
\left[I_{\widehat{W}_{25}}^{\rm lqc}\right]_{K\bar{K}2\p}&=&0.26
(26)\times 10^{-3}
\ ,\nonumber
\end{eqnarray}
and
\begin{eqnarray}
\label{KKbarpipispd}
\left[a_\m^{W1,\rm s+lqd}\right]_{K\bar{K}2\p}&=&0.58(60)
\times 10^{-10}
\ ,\\
\left[a_\m^{W2,\rm s+lqd}\right]_{K\bar{K}2\p}&=&0.0032(31)
\times 10^{-10}
\ ,\nonumber\\
\left[I_{\widehat{W}_{15}}^{\rm s+lqd}\right]_{K\bar{K}2\p}&=&-0.027
(28)\times 10^{-2}
\ ,\nonumber\\
\left[I_{\widehat{W}_{25}}^{\rm s+lqd}\right]_{K\bar{K}2\p}&=&0.25
(26)\times 10^{-3}
\ .\nonumber
\end{eqnarray}

The final G-parity-ambiguous modes for which additional external
experimental input provides an improved isospin decomposition are the
two radiative modes $\pi^0\gamma$ and $\eta\gamma$. Using the accurate
VMD representations of the $e^+ e^-\rightarrow \pi^0\gamma$ and
$e^+e^-\rightarrow \eta \gamma$ cross sections detailed in
Appendix~\ref{PgammaVMDdecomp}, and employing 2023 PDG input 
\cite{Workman:2022ynf}, we find the
following results for the lqc and s+lqd contributions from these modes:
\begin{eqnarray}
\label{pi0gammapetagammalqc}
\left[a_\m^{W1,\rm lqc}\right]_{\pi^0\gamma +\eta\gamma}&=&
\left( 0.074(13)+0.063(0)\right) \times 10^{-10}=0.137(13)\times 10^{-10}
\ ,\\
\left[a_\m^{W2,\rm lqc}\right]_{\pi^0\gamma +\eta\gamma}&=&
\left( 0.029(5)+0.021(0)\right) \times 10^{-10}=0.050(5)\times 10^{-10}
\ ,\nonumber\\
\left[I_{\widehat{W}_{15}}^{\rm lqc}\right]_{\pi^0\gamma +\eta\gamma}&=&
\left( 0.018(3)+0.013(0)\right) \times 10^{-2}=0.031(3)\times 10^{-2}
\ ,\nonumber\\
\left[I_{\widehat{W}_{25}}^{\rm lqc}\right]_{\pi^0\gamma +\eta\gamma}&=&
\left( 0.031(6)+0.028(0)\right) \times 10^{-3}=0.059(6)\times 10^{-3}
\ ,\nonumber
\end{eqnarray}
and
\begin{eqnarray}
\label{pi0gammapetagammaslqd}
\left[a_\m^{W1,\rm s+lqd}\right]_{\pi^0\gamma +\eta\gamma}&=&
\left( 1.25(6)+0.25(1)\right) \times 10^{-10}\, =\, 1.50(6)\times 10^{-10}
\ ,\\
\left[a_\m^{W2,\rm s+lqd}\right]_{\pi^0\gamma +\eta\gamma}&=&
\left( 0.54(3)+0.04(0)\right) \times 10^{-10}\, =\, 0.58(3)\times 10^{-10}
\ ,\nonumber\\
\left[I_{\widehat{W}_{15}}^{\rm s+lqd}\right]_{\pi^0\gamma +\eta\gamma}&=&
\left( 0.32(2)+0.03(0)\right) \times 10^{-2}\, =\, 0.35(2)\times 10^{-2}
\ ,\nonumber\\
\left[I_{\widehat{W}_{25}}^{\rm s+lqd}\right]_{\pi^0\gamma +\eta\gamma}&=&
\left( 0.53(3)+0.11(0)\right) \times 10^{-3}\, =\, 0.64(3)\times 10^{-3}
\ ,\nonumber
\end{eqnarray}
where the first term in each intermediate expression is the
$\pi^0\gamma$ contribution and the second term the $\eta\gamma$ one.

Contributions from the remaining $G$-parity-ambiguous modes,
$K\bar{K}3\p$, $\omega(\rightarrow npp)K\bar{K}$,
$\eta (\rightarrow npp) K\bar{K}$ (no $\phi$), $p\bar{p}$, $n\bar{n}$,
and low-$s$ $\pi^0 \gamma$ and $\eta \gamma$, which are very small,
are listed for completeness in Appendix~\ref{otherweights}.

The sums of all exclusive-mode contributions below $s=(1.937)^2$~GeV$^2$
for the lqc window quantities are obtained from
Eqs.~(\ref{Gparlqc}),~(\ref{KKbarlqc}),~(\ref{KKbarpilqc}),~(\ref{KKbarpipilqc}),
~(\ref{pi0gammapetagammalqc}) and the further exclusive-mode contributions
listed in Appendix~\ref{otherweights}. The results are 
\begin{eqnarray}
\label{excllqc}
\left[a_\m^{W1,\rm lqc}\right]_{\rm excl}&=&188.82(1.01)\times 10^{-10}
\ ,\\
\left[a_\m^{W2,\rm lqc}\right]_{\rm excl}&=&94.60(36)\times 10^{-10}
\ ,\nonumber\\
\left[I_{\widehat{W}_{15}}^{\rm lqc}\right]_{\rm excl}&=&43.75(16)
\times 10^{-2}
\ ,\nonumber\\
\left[I_{\widehat{W}_{25}}^{\rm lqc}\right]_{\rm excl}&=&75.59(43)
\times 10^{-3}
\ .\nonumber
\end{eqnarray}
For the s+lqd window quantities, similarly, we obtain from Eqs.~(\ref{Gparspd}),
~(\ref{KKbarspd}),~(\ref{KKbarpispd}),~(\ref{KKbarpipispd}),
~(\ref{pi0gammapetagammaslqd}) and the further exclusive-mode contributions
listed in Appendix~\ref{otherweights}, the results 
\begin{eqnarray}
\label{exclspd}
\left[a_\m^{W1,\rm s+lqd}\right]_{\rm excl}&=&23.85(74)
\times 10^{-10}
\ ,\\
\left[a_\m^{W2,\rm s+lqd}\right]_{\rm excl}&=&-0.04(14)
\ ,\nonumber\\
\left[I_{\widehat{W}_{15}}^{\rm s+lqd}\right]_{\rm excl}&=&1.875(92)
\times 10^{-2}
\ ,\nonumber\\
\left[I_{\widehat{W}_{25}}^{\rm s+lqd}\right]_{\rm excl}&=&11.26(32)
\times 10^{-3}
\ .\nonumber
\end{eqnarray}

\begin{boldmath}
\subsection{\label{pertth}Perturbative contribution above $s=(1.937)^2$~GeV$^2$}
\end{boldmath}
Above $s=(1.937)^2$~GeV$^2$ we use perturbation theory to evaluate
(three-flavor) contributions to the window quantities. We employ massless
perturbation theory,\footnote{Mass corrections are negligibly small, see
Ref.~\cite{spd}.} using the five-loop result of Ref.~\cite{Baikov:2008jh} for the
Adler function, following the steps outlined in Refs.~\cite{lqc,spd}. Above
$s=(1.937)^2$~GeV$^2$ and up to the charm-quark threshold, perturbation
theory agrees well with inclusive BES~\cite{BES:2001ckj,BES:2009ejh} and
KEDR~\cite{KEDR:2018hhr} $R(s)$ measurements, although the agreement with
recent, more precise, BESIII results~\cite{ BESIII:2021wib} is less good.
We find for these perturbative contributions, using
$\a_s(m_\t^2)=0.3139(71)$ \cite{Workman:2022ynf}
\begin{eqnarray}
\label{pertthlqc}
\left[a_\m^{W1,\rm lqc}\right]_{\rm pert.th.}&=&10.896\times 10^{-10}
\ ,\\
\left[a_\m^{W2,\rm lqc}\right]_{\rm pert.th.}&=&0.00757\times 10^{-10}
\ ,\nonumber\\
\left[I_{\widehat{W}_{15}}^{\rm lqc}\right]_{\rm pert.th.}&=&-0.574
\times 10^{-2}
\ ,\nonumber\\
\left[I_{\widehat{W}_{25}}^{\rm lqc}\right]_{\rm pert.th.}&=&3.658
\times 10^{-3}
\ .\nonumber
\end{eqnarray}
Error estimates along the lines of Ref.~\cite{spd} lead to errors of the order
of 0.1\% of the central values in Eq.~(\ref{pertthlqc}), which are small enough
that they can be ignored in our final results. In order to obtain the
s+lqd perturbative contributions, the values in Eq.~(\ref{pertthlqc}) have to
be multiplied by $(2/9)(9/10)=1/5$ \cite{lqc}. For completeness, we
provide the resulting values:
\begin{eqnarray}
\label{pertthspd}
\left[a_\m^{W1,\rm s+lqd}\right]_{\rm pert.th.}&=&2.179\times 10^{-10}
\ ,\\
\left[a_\m^{W2,\rm s+lqd}\right]_{\rm pert.th.}&=&0.00151\times 10^{-10}
\ ,\nonumber\\
\left[I_{\widehat{W}_{15}}^{\rm s+lqd}\right]_{\rm pert.th.}&=&-0.115
\times 10^{-2}
\ ,\nonumber\\
\left[I_{\widehat{W}_{25}}^{\rm s+lqd}\right]_{\rm pert.th.}&=&0.732
\times 10^{-3}
\ .\nonumber
\end{eqnarray}

Although the use of perturbation theory above $s=(1.937)^2$~GeV$^2$ is
supported by the good agreement with the inclusive $R(s)$ data of
Refs.~\cite{BES:2001ckj,BES:2009ejh,KEDR:2018hhr}, the slight tension
with the recent BESIII data~\cite{ BESIII:2021wib} hints at possible
residual duality violations (DVs) even in the inclusive region. While
our estimates for residual DV contributions to $a_\m^{\rm HVP,lqc}$ and
$a_\mu^{\rm HVP,s+lqd}$ showed these to be small, DVs represent an
intrinsic limitation of perturbation theory. We will thus include
the central values of our DV estimates in our final results, assigning
them an uncertainty of 100\%. The assigned DV uncertainty totally
dominates our estimate of the uncertainty associated with the use
of perturbation theory in the inclusive region.

For lqc contributions, which involve only the $\rho_{\rm EM}^{I=1}(s)$
spectral function, we obtain our DV estimates using the results of
finite-energy sum-rule fits performed in Ref.~\cite{Boito:2020xli}
using an improved version of the $I=1$ charged-current spectral
function, $\rho_{ud;V}(s)$, obtained mainly from $\tau$ decay data, and
related by CVC to $\rho_{\rm EM}^{I=1}(s)$ by
$\rho_{\rm EM}^{I=1}(s)=\frac{1}{2}\rho_{ud;V}(s)$. Following the
procedure described in Sec. IIIA of Ref.~\cite{lqc}, we find for our
estimates of the DV contributions to $a_\mu^{W, {\rm lqc}}$ and
$I_W^{\rm lqc}$ the results
\begin{eqnarray}
\label{DVs-lqc}
\left[a_\m^{W1,\rm lqc}\right]_{\rm DVs}&=&0.162(77)\times 10^{-10}
\ ,\\
\left[a_\m^{W2,\rm lqc}\right]_{\rm DVs}&=&0.00034(20)\times 10^{-10}
\ ,\nonumber\\
\left[I_{\widehat{W}_{15}}^{\rm lqc}\right]_{\rm DVs}&=&-0.0092(44)
\times 10^{-2}
\ ,\nonumber\\
\left[I_{\widehat{W}_{25}}^{\rm lqc}\right]_{\rm DVs}&=&0.069(33)
\times 10^{-3}
\ .\nonumber
\end{eqnarray}
Possible residual DV corrections to the use of perturbation theory
in the inclusive region will be present in both the $I=1$ and $I=0$
contributions to $a_\mu^{W, {\rm s+lqd}}$ and $I_W^{\rm s+lqd}$.
To estimate the combination of these effects, we will use the results of
the finite-energy sum-rule fits to KNT $\rho_{\rm EM}(s)$ data performed
in Ref.~\cite{Boito:2018yvl}. We find
\begin{eqnarray}
\label{DVs-s+lqd}
\left[a_\m^{W1,\rm s+lqd}\right]_{\rm DVs}&=&-0.17(5)\times 10^{-10}
\ ,\\
\left[a_\m^{W2,\rm s+lqd}\right]_{\rm DVs}&=&-0.0004(1)\times 10^{-10}
\ ,\nonumber\\
\left[I_{\widehat{W}_{15}}^{\rm s+lqd}\right]_{\rm DVs}&=&
0.010(3)\times 10^{-2}
\ ,\nonumber\\
\left[I_{\widehat{W}_{25}}^{\rm s+lqd}\right]_{\rm DVs}&=&
-0.07(2)\times 10^{-3}
\ .\nonumber
\end{eqnarray}

The results of Eqs.~(\ref{pertthlqc}) and~(\ref{pertthspd}), supplemented
by the central values from Eqs.~(\ref{DVs-lqc}) and Eqs.~(\ref{DVs-s+lqd}),
which serve as our estimates for the DV-induced perturbative uncertainties,
have to be added to those of Eqs.~(\ref{excllqc}) and~(\ref{exclspd}), respectively.
As noted above, we take the total error on the perturbative plus DV
contributions to be equal to the (absolute value of the) central values
of the DV contributions,
which are always larger than the DV errors quoted above.
This produces the following total lqc and s+lqd
contributions, not yet corrected for EM and SIB effects:
\begin{eqnarray}
\label{totallqc}
a_\m^{W1,\rm lqc}&=&199.88(1.02)\times 10^{-10}
\ ,\\
a_\m^{W2,\rm lqc}&=&94.61(36)\times 10^{-10}
\ ,\nonumber\\
I_{\widehat{W}_{15}}^{\rm lqc}&=&43.17(16)\times 10^{-2}
\ ,\nonumber\\
I_{\widehat{W}_{25}}^{\rm lqc}&=&79.32(43)\times 10^{-3}
\ ,\nonumber
\end{eqnarray}
and
\begin{eqnarray}
\label{totalspd}
a_\m^{W1,\rm s+lqd}&=&25.86(76)\times 10^{-10}
\ ,\\
a_\m^{W2,\rm s+lqd}&=&-0.04(14)\times 10^{-10}
\ ,\nonumber\\
I_{\widehat{W}_{15}}^{\rm s+lqd}&=&1.770(92)\times 10^{-2}
\ ,\nonumber\\
I_{\widehat{W}_{25}}^{\rm s+lqd}&=&11.92(32)\times 10^{-3}
\ .\nonumber
\end{eqnarray}

\vskip0.8cm
\section{\label{EMSIB} Electromagnetic and strong isospin-breaking effects}
The results of Eqs.~(\ref{totallqc}) and ~(\ref{totalspd}) were obtained using
experimental data and thus contain EM and SIB effects. These effects need
to be estimated and subtracted to obtain lqc and s+lqd results that can be
compared directly to those obtained from isospin-symmetric lattice QCD
without QED. We employ the same strategy to carry out these subtractions
as that used in Refs.~\cite{lqc,spd}, which is predicated on the
observation that, to first order in IB, SIB contributions occur only in
the MI component of $\rho_{\rm EM}(s)$, while EM contributions are present
in all of the pure $I=1$, pure $I=0$ and MI components.
IB corrections to the window quantities we consider are thus of two
types. Those present in the pure $I=1$ and pure $I=0$ contributions are, to
first order in IB, purely EM, and require only an estimate of the inclusive
combination of EM contributions from all exclusive modes, with no need
for a further breakdown of these corrections into those associated with
individual exclusive modes. The MI corrections, in contrast, require
removing from the nominal pure $I=1$ and pure $I=0$ sums obtained above
exclusive-mode IB contributions which, in fact, represent MI contaminations
of those sums. Examples of such MI contaminations are the two-pion and
three-pion contributions resulting from the $\rho-\omega$-mixing-induced
processes $e^+e^-\rightarrow \omega\rightarrow \rho\rightarrow 2\pi$ and
$e^+e^-\rightarrow \rho\rightarrow \omega\rightarrow 3\pi$. MI corrections
to the lqc and s+lqd combinations thus require identifying the combined
EM+SIB IB parts of the individual exclusive-mode contributions relevant
to each, and cannot be carried out using inclusive versions of the
EM or SIB contributions, or their sum. We do, however, expect the MI
contaminations to be dominated by contributions from the two- and
three-pion modes, where the narrowness of the $\omega$ peak and the
small $\rho-\omega$ mass difference lead to a strong enhancement of
IB contributions from the $\rho-\omega$ region. We will estimate MI
two- and three-pion corrections from this region using fits to data
(which, of course, produce assessments of the EM+SIB sum) and employ
a generic $O(1\%)$ estimate for the size of MI contamination in the
contributions from other exclusive modes, which (i) lie higher in the
spectrum, and hence have contributions suppressed by the falloff in $s$
of the weights considered here, and (ii) are not subject to any
narrow-nearby-resonance enhancements. Given the very small size of EM
and SIB corrections to the perturbative contribution to the lqc and
s+lqd spectral functions, we will ignore IB corrections from the
$s\ge (1.937)^2\ {\rm GeV}^2$ (inclusive) region.

In view of the above discussion, we treat separately the corrections
for EM effects in the nominally pure $I=1$ and $I=0$ contributions
and those for EM+SIB MI contamination, discussing the former in Sec.~\ref{EM}
and the latter in Sec.~\ref{SIB}. As mentioned above, we follow many lattice
groups and define our isospin limit of QCD as one in which all pions
have a
 mass equal to the physical neutral pion mass.

\begin{boldmath}
\subsection{\label{EM} $I=1$ and $I=0$ electromagnetic corrections}
\end{boldmath}
To quantify and subtract the EM contributions present in the
pure $I=1$ and $I=0$ parts of the window quantities of interest,
we rely on information from the lattice results obtained in
Ref.~\cite{BMW}. While some EM effects have been estimated from experimental
data \cite{Colangelo:2022prz,Hoferichter:2023bjm,Hoferichter:2023sli}, additional potentially significant EM
effects have not\footnote{For an expanded discussion of this point
see, {\it e.g.,} the appendix of Ref.~\cite{lqc}.} and we thus consider it
unavoidable to rely on lattice EM data for the EM corrections. The
existence of significant cancellations amongst the set of data-based EM
contribution estimates detailed in Refs.~\cite{Hoferichter:2022iqe,Hoferichter:2023sli} also
argues in favor of using the inclusive lattice result since that
result necessarily includes contributions from all sources, including
those not currently amenable to data-based estimates, and whose
relative role might be enhanced by the strong cancellations amongst
the currently quantified contributions. This constitutes our only
use of lattice data in obtaining our estimates for the lqc and s+lqd
window quantities; as we will see, these corrections are very small.

For the RBC/UKQCD intermediate window $W1$, the lqc EM correction has been
obtained directly in Ref.~\cite{BMW}. The result is
\begin{equation}
\label{W1EMlqc}
\D_{\rm EM}a_\m^{W1,\rm lqc}=0.035(59)\times 10^{-10}\ .
\end{equation}
For the s+lqd EM correction for the same window, the relevant lattice data
are also given in Ref.~\cite{BMW}. Using exactly the same strategy as in
Ref.~\cite{spd}, we find
\begin{equation}
\label{W1EMspd}
\D_{\rm EM}a_\m^{W1,\rm s+lqd}=0.012(11)\times 10^{-10}\ .
\end{equation}
As Ref.~\cite{BMW} did not obtain the relevant lattice estimates for
EM contributions to the other window quantities considered here,
we do not have equivalent estimates for these other quantities.
However, if we compare the $W1$ corrections with Eqs.~(\ref{totallqc})
and~(\ref{totalspd}), we see that the central value of the lqc EM correction
is $\sim$30 times smaller than the error in Eqs.~(\ref{totallqc}), while the
central value of the s+lqd correction is $\sim$60 times smaller than the
error in Eqs.~(\ref{totalspd}). Since the relative errors on the other
window quantities in Eqs.~(\ref{totallqc}) and~(\ref{totalspd}) are of
order the same size as or larger than those in the $W1$ quantities,
we will assume that EM corrections to the pure $I=1$ and
$I=0$ contributions to these quantities can be safely neglected,
and ignore these EM corrections in the rest of this paper.
This assumption should be particularly safe for the s+lqd contributions,
where the diagrammatic analysis of Ref.~\cite{spd} shows the existence of
generic strong cancellations, for example in the numerically dominant
light-quark EM valence-valence connected and disconnected contributions.

%\newpage
\subsection{\label{SIB} The mixed-isospin (MI) correction}
As in Ref.~\cite{spd}, we expect the MI EM+SIB correction to be dominated by
contributions from the 2-pion and 3-pion exclusive modes, where there
are potentially strong enhancements due to $\r-\o$ interference
in the $\rho-\omega$ resonance region, and where contributions
from that region are more strongly weighted than are those of other
modes, lying at higher $s$, in all the window quantities considered
in this paper. Such IB $\r-\o$ region $2\pi$ and $3\pi$ contributions
can be estimated from the interference terms in fits to the
$e^+e^-\rightarrow 2\p$ and $e^+e^-\rightarrow 3\p$ electroproduction
cross sections associated with the $\rho-\omega$-mixing-induced IB
processes $e^+e^-\rightarrow \omega\rightarrow\rho\rightarrow 2\pi$
and $e^+e^-\rightarrow\rho\rightarrow\omega\rightarrow 3\pi$, which,
to first order in $\a$ and $m_u-m_d$, produce contributions
lying entirely in the MI contribution, $\r_{\rm EM}^{\rm MI}$,
of Eq.~(\ref{Idecomp}).

For the window quantity $a_\m^{W1}$, the $\r-\o$-mixing-enhanced,
MI two-pion exclusive-mode contribution has been obtained in
Refs.~\cite{Colangelo:2022prz,Hoferichter:2023sli} from fits to two-pion electroproduction data
employing a dispersively constrained representation of the
pion form factor incorporating the effect of $\rho-\omega$
mixing.\footnote{The fits of Refs.~\cite{Colangelo:2022prz,Hoferichter:2023sli}, of course,
also provide determinations of the $\r$-$\o$-mixing-enhanced, MI
two-pion exclusive-mode contribution to $a_\mu^{\rm HVP}$. The result
of the most recent update \cite{Hoferichter:2023sli}
in that case is $3.79(19)\times 10^{-10}$.} The result,
\begin{equation}
\label{W1MIdisp}
\left[ a_\m^{W1}\right]_{\pi\pi}^{\rm MI}=0.86(6)\times 10^{-10}\ ,
\end{equation}
represents a MI contamination to be subtracted from the nominal $I=1$
exclusive-mode sum, and hence produces a correction
\begin{equation}
\label{W1MI2pilqccorrn}
\Delta_{\pi\pi}^{\rm MI}a_\mu^{W1, {\rm lqc}}=
-{\frac{10}{9}} \times 0.86(6)
\times 10^{-10} = -0.96(7)\times 10^{-10}
\end{equation}
to the nominal $a_\mu^{W1, {\rm lqc}}$ result of Eqs.~(\ref{totallqc}). In spite
of its enhancement by $\rho -\omega$ mixing, this correction is only
$\sim 0.6\%$ of the total two-pion contribution to $a_\mu^{W1}$. We thus
consider it extremely safe to assume that the magnitude of the sum of MI
corrections to the nominal $I=1$ sum from other exclusive  modes having
no analogous narrow-nearby-resonance enhancements is less than $1\%$
of the sum, $25.68\times 10^{-10}$, of contributions to $a_\mu^{W1}$
from those modes. We thus add a further uncertainty
$\pm (10/9)\times 0.26\times 10^{-10} =\pm 0.29\times 10^{-10}$ to
that in Eq.~(\ref{W1MI2pilqccorrn}) and take as our final estimate for
the MI correction to the pre-IB-corrected $a_\mu^{W1, {\rm lqc}}$
value of Eq.~(\ref{totallqc}), the result
\begin{equation}
\label{W1MItotlqccorrn}
\Delta^{\rm MI}a_\mu^{W1, {\rm lqc}}= -0.96(7)(29)\times 10^{-10}\ .
\end{equation}
This result is compatible within errors with the expectation,
$-0.753(43)\times 10^{-10}$, for the SIB part of the MI lqc correction
implied by the lattice result of Ref.~\cite{BMW} for the SIB contribution to
$a_\mu^{W1,{\rm lqc}}$, and hence with the expected dominance of the
MI EM+SIB lqc correction by its SIB component.
Neglecting the very small pure $I=1$ EM correction of
Eq.~(\ref{W1EMlqc}), we obtain the IB-corrected result
\begin{eqnarray}
\label{finallqc}
a_\m^{W1,\rm lqc}&=&\left(199.88(1.02)-0.96(0.30)\right)
\times 10^{-10} =198.9(1.1)\times 10^{-10}\ .
\end{eqnarray}

Two-pion MI IB corrections due to $\r-\o$ mixing based on the
analysis of Ref.~\cite{Colangelo:2022prz,Hoferichter:2023sli} have also been made available
to us for the other three window quantities \cite{HS} considered here.
In these cases, there are no lattice results to compare with. The
resulting MI contributions to the nominal $I=1$ sums are
$0.767(31)\times 10^{-10}$ for the  $W2$ window quantity, and
$0.331(13)\times 10^{-2}$ and $0.300(20)\times 10^{-3}$ for the
$\widehat{W}_{15}$ and $\widehat{W}_{25}$ window quantities. To
obtain the corresponding lqc IB corrections, these need to be
multiplied by 10/9, and subtracted from the totals in
Eq.~(\ref{totallqc}). The non-$2\p$ exclusive-mode contributions
to these quantities are $0.9\times 10^{-10}$, vanishingly small, and
$13\times 10^{-3}$ respectively. Taking, as above, 1\% of these
contributions as a further uncertainty induced by MI IB effects from
nominally $I=1$, non-$2\pi$ exclusive modes, leads to our final
estimates
\begin{eqnarray}
\label{finallqcother}
a_\m^{W2,\rm lqc}&=&\left(94.61(36)-\frac{10}{9}\times
0.767(32)\right)\times 10^{-10}
=93.75(36)\times 10^{-10}\ ,\nonumber\\
I_{\widehat{W}_{15}}^{\rm lqc}&=&\left(43.17(16)-\frac{10}{9}\times
0.331(13)\right)\times 10^{-2}
=42.80(16)\times 10^{-2}\ ,\nonumber\\
I_{\widehat{W}_{25}}^{\rm lqc}&=&\left(79.32(43)-\frac{10}{9}\times
0.30(13)\right)\times 10^{-3}
=78.99(45)\times 10^{-3}\,.
\end{eqnarray}

Unlike estimates for the MI corrections for the lqc components of the
window quantities considered in this paper, which, as explained above,
can be obtained using as key input the results of the dispersively
constrained analysis of experimental $2\p$ electroproduction data
detailed in Ref.~\cite{Colangelo:2022prz}, estimates of the MI corrections
for the analogous s+lqd components require also input on what is
expected to be the dominant nominally $I=0$ MI correction, namely
that from the $3\p$ exclusive mode. While the fact that $2\p$
contributions to the window quantities listed in Tables \ref{tab1},
\ref{tab4}, \ref{tab5} and \ref{tab6} exceed the corresponding $3\p$
contributions by factors of $7.0-13.6$, might lead one to expect the
$3\p$ MI corrections to be similarly smaller than the corresponding
$2\p$ MI ones, this is, in fact, unlikely to be the case due to an
countervailing numerical enhancement of the relative
$\rho -\omega$-mixing-induced $3\p$ correction.

The existence of this enhancement can be understood as follows. The ratio of
the $\rho -\omega$-mixing-induced IB interference contribution to the
dominant isospin-conserving (IC) $\rho$ contribution to the
$\rho -\omega$-resonance-region $2\p$ cross sections is proportional to
the product $P_{2\pi}\equiv \epsilon_{\rho\omega}f_\omega /f_\rho$, with
$\epsilon_{\rho\omega}$ the parameter characterizing the strength of
$\rho -\omega$ mixing and $f_V$, $V=\rho ,\, \omega$ the vector-meson
decay constants, characterizing the strength of the couplings of the $\rho$
and $\omega$ to the EM current. The analogous ratio, of
$\rho -\omega$-mixing-induced-IB-interference to $\omega$-dominated
IC contributions to the resonance-region 3-pion cross sections,
is, in contrast, proportional to the product
$P_{3\pi}\equiv\epsilon_{\rho\omega}f_\rho /f_\omega$. Experimentally
(as expected for near-ideal mixing of the vector meson nonet),
$f_\rho \simeq 3f_\omega$. A natural enhancement, by a factor of
$P_{3\pi}/P_{2\pi}\simeq 9$, is thus present in the ratio of relative
sizes of $\rho -\omega$-mixing-induced resonance-region
IB in the $3\p$ versus $2\p$ channels. It is thus unlikely
that the MI $3\p$ correction is safely negligible on the scale
of the MI $2\p$ correction.

The $\r-\o$-mixing MI corrections in the $3\p$ channel, relevant for the
$I=0$ contribution and thus the s+lqd contribution, have recently been
estimated in Refs.~\cite{Hoferichter:2023bjm,Hoferichter:2023sli}, for
$a_\mu^{\rm HVP}$ and $a_\mu^{W1}$. For $a_\mu^{W1}$, their estimate,
which is nominally a contribution to the $I=0$ part of $a_\m^{W1}$, is
\begin{equation}
\label{MI3pizW1}
\left[a_\m^{W1}\right]_{3\p}^{\rm MI}=-1.03(27)\times 10^{-10} .
\end{equation}
From Eqs.~(\ref{excllqc}) and~(\ref{exclspd}) and using Table \ref{tab1} one
finds that the other-than $3\p$ exclusive-mode contribution for $I=0$
equals $22.53\times 10^{-10}$, 1\% of which is $0.23$. We
add this as an additional error (in quadrature) to Eq.~(\ref{MI3pizW1}), arriving
at $-1.03(35)\times 10^{-10}$. To find the MI correction to the s+lqd part
of $a_\m^{W1}$, we need to subtract this, while adding back in $1/10$
times Eq.~(\ref{W1MItotlqccorrn}) (\seef\ Eq.~(\ref{spdspec})). This leads to
our estimate for the MI correction
\begin{equation}
\label{MIsplqd}
\Delta^{\rm MI}a_\mu^{W1, {\rm s+lqd}}=1.13(36)\times 10^{-10}\ ,
\end{equation}
where we combined errors in Eqs.~(\ref{MI3pizW1}) and~(\ref{W1MItotlqccorrn})
ignoring correlations. We thus obtain the IB-corrected result\footnote{We
again neglect EM corrections, which are very small.}
\begin{eqnarray}
\label{finalslqd}
a_\m^{W1,\rm s+lqd}&=&\left(25.86(76)+1.13(36)\right)
\times 10^{-10} =27.0(8)\times 10^{-10}\ .
\end{eqnarray}

Three-pion MI IB corrections due to $\r-\o$ mixing based on the analysis
of Refs.~\cite{Hoferichter:2023bjm,Hoferichter:2023sli} have also been made
available to us for the other three window quantities \cite{HS} considered
here. The resulting MI contributions to the nominal $I=0$ sums are
$-0.367(101)\times 10^{-10}$ for the $W2$ window quantity, and
$-0.230(61)\times 10^{-2}$ and $-0.44(12)\times 10^{-3}$ for the
$\widehat{W}_{15}$ and $\widehat{W}_{25}$ window quantities.
The non-$3\p$ exclusive-mode contributions to these quantities are
$2.7\times 10^{-10}$, $2.2\times 10^{-2}$ and
$10.1\times 10^{-3}$, respectively.\footnote{To be conservative,
we take the absolute value of all contributions in Table~\ref{tab5}.}
Taking, as above, 1\% of these contributions as a further uncertainty
induced by MI IB effects from nominally $I=0$, non-$3\pi$
exclusive modes, leads to our final estimates 
\begin{eqnarray}
\label{finalslqdother}
a_\m^{W2,\rm s+lqd}&=&\left(-0.03(14)+0.37(11)
+\frac{1}{9}\times 0.767(33)\right)\times 10^{-10}\\
&=&0.42(18)\times 10^{-10}\ ,\nonumber\\
I_{\widehat{W}_{15}}^{\rm s+lqd}&=&\left(1.770(92)+
0.230(65)+\frac{1}{9}\times 0.331(13)\right)\times 10^{-2}\nonumber\\
&=&2.04(11)\times 10^{-2}\ ,\nonumber\\
I_{\widehat{W}_{25}}^{\rm s+lqd}&=&\left(11.92(32)+
0.44(16)+\frac{1}{9}\times 0.30(13)\right)\times 10^{-3}\nonumber\\
&=&12.39(36)\times 10^{-3}\,.\nonumber
\end{eqnarray}
The terms between parentheses come from Eq.~(\ref{totalspd}), and the
results for the MI $I=0$ $3\pi$ and MI $I=1$ $2\pi$ corrections quoted
above, respectively.

It is possible to test the treatment of exclusive-mode MI
contributions described above by comparing the inclusive MI sums
that treatment implies to the corresponding inclusive MI lattice
results, available for $a_\mu^{\rm HVP}$ and $a_\mu^{W1}$ from
Ref.~\cite{BMW}. The MI lattice results are obtained by combining the
SIB results from Ref.~\cite{BMW} with EM MI estimates obtained using
the EM results quoted in Ref.~\cite{BMW}, following the diagrammatically
based analysis strategy employed in Ref.~\cite{spd}. The latter analysis
yields the result, $-0.49(25)\times 10^{-10}$, obtained already in
Ref.~\cite{spd}, for the inclusive MI EM contribution to $a_\mu^{\rm HVP}$,
and $-0.022(23)\times 10^{-10}$ for the MI EM contribution to $a_\mu^{W1}$.
For the SIB contributions, which, to first order in IB, are pure MI, we
have, summing the quoted connected and disconnected contributions, the
lattice results \cite{BMW}
\begin{eqnarray}
\label{bmwsibresults}
\left[ a_\mu^{\rm HVP}\right]_{\rm SIB}&=&1.93(1.20)\times 10^{-10}\ ,\\
\left[ a_\mu^{W1}\right]_{\rm SIB}&=&0.516(44)\times 10^{-10}\ .\nonumber
\end{eqnarray}
The inclusive MI lattice totals are thus
\begin{eqnarray}
\label{bmwMIinclresults}
\left[ a_\mu^{\rm HVP}\right]_{\rm MI, latt}&=&1.44(1.23)\times 10^{-10}\
,\\
\left[ a_\mu^{W1}\right]_{\rm MI, latt}&=&0.494(50)\times 10^{-10}\ .
\nonumber
\end{eqnarray}
In the treatment above, the inclusive MI total is, in contrast, obtained
by summing our estimates for the MI $\pi^0\gamma$, $\eta \gamma$, $2\pi$
and $3\pi$ exclusive-mode contributions, with an additional uncertainty
equal to $1\%$ of the sums of the contributions for all other exclusive
modes. The $2\pi$ and $3\pi$ contributions are those detailed above,
while the MI $\pi^0\gamma$ and $\eta \gamma$ contributions are obtained
using the same VMD model used to determine the corresponding $I=0$ and $I=1$
contributions, outlined in Appendix~\ref{PgammaVMDdecomp}. The results
for the latter are
\begin{eqnarray}
&&\left[ a_\mu^{\rm HVP}\right]^{\rm MI}_{\pi^0\gamma +\eta\gamma}
=\left( 0.733(65)+0.067(24)\right)\times 10^{-10} = 0.800(69)\times 10^{-10}
\nonumber\\
&&\left[ a_\mu^{W1}\right]^{\rm MI}_{\pi^0\gamma +\eta\gamma}
=\left( 0.268(24)+0.026(1)\right)\times 10^{-10} = 0.294(24)\times 10^{-10}\, ,
\label{pi0gammaetagammaMiintegrals}\end{eqnarray}
where the first terms is the $\pi^0\gamma$ contribution and the second
the $\eta\gamma$ contribution. Combining these results with those from
the other modes, we find the following alternate data-driven (dd) results
\begin{eqnarray}
\label{datadrivenMIinclresults}
\left[ a_\mu^{\rm HVP}\right]_{\rm MI, dd}&=&1.91(73)(77)\times 10^{-10}\
,\\
\left[ a_\mu^{W1}\right]_{\rm MI, dd}&=&0.12(28)(49)\times 10^{-10}\
,\nonumber
\end{eqnarray}
where the first error is the quadrature sum of the errors on the MI
$\pi^0\gamma$, $\eta \gamma$, $2\pi$ and $3\pi$ contributions and the
second is our estimate of the uncertainty produced by neglecting MI
contributions from all other exclusive modes. The data-driven estimates
are compatible within errors with the lattice results in both cases.

\begin{boldmath}
\section{\label{updatesprevious}Updates of our previous determinations of $a_\mu^{\rm HVP, s+lqd}$,
$a_\mu^{\rm HVP,lqc}$ and $a_\mu^{\rm W1,lqc}$}
\end{boldmath}
In Refs.~\cite{lqc,spd,Benton:2023dci} we provided first data-driven estimates
for $a_\mu^{\rm HVP,lqc}$, $a_\mu^{\rm HVP, s+lqd}$ and $a_\mu^{\rm W1,lqc}$,
respectively. This section updates the results of those analyses,
taking into account (i) an improved treatment of the small contributions from
the $\pi^0\gamma$ and $\eta\gamma$ exclusive modes and (ii) changes in
external input for the MI $2\pi$ and $3\pi$ IB corrections. In those
earlier analyses, the exclusive-mode $\pi^0\gamma$ and $\eta \gamma$
contributions were, based on the dominance of the experimental cross
sections by the large $\omega$ and $\phi$ peaks, assigned to the nominally
pure $I=0$ category. The VMD representation of those cross sections,
outlined in Appendix~\ref{PgammaVMDdecomp}, allows for an improved version
of this treatment.{\footnote{We thank Martin Hoferichter for bringing this
possibility to our attention.}} Because the resulting pure $I=1$
contributions are very small, this improvement has only a small effect on
the previous lqc results. It has a larger (though still small) impact
on the $a_\mu^{\rm HVP,s+lqd}$ result. There have also been two small
shifts in the input for the MI $2\pi$ contribution since the preliminary
version of the Ref.~\cite{Colangelo:2022prz} HVP result used in the
$a_\mu^{\rm HVP, s+lqd}$ determination of Ref.~\cite{spd}. The second of
these shifts also affects the $a_\mu^{\rm HVP,lqc}$ and $a_\mu^{\rm W1,lqc}$
results of Refs.~\cite{lqc,Benton:2023dci}. The impacts of these shifts on the
lqc results are very small on the scale of the errors on those previous
results. Finally, the results of Ref.~\cite{Hoferichter:2023bjm} for the MI
$3\pi$ corrections provide a significant improvement to the earlier
treatment of those corrections and hence to the reliability of the
determination of $a_\mu^{\rm HVP, s+lqd}$. The numerical impacts of these
changes are quantified below.

Equations~(\ref{pi0gammapetagammalqc}) and~(\ref{pi0gammapetagammaslqd})
provide the updated versions of the lqc and s+lqd contributions from the
$\pi^0\gamma$ and $\eta\gamma$ modes. The contributions of these modes to
$a_\mu^{\rm HVP,lqc}$ and $a_\mu^{\rm W1,lqc}$, which were previously
taken to be zero, are now $0.36(4)\times 10^{-10}$ and $0.14(1)\times
10^{-10}$, respectively. The updated version of the contribution to
$a_\mu^{\rm HVP,s+lqd}$ is $4.00(17)\times 10^{-10}$, representing
shifts of $-1.16(14)\times 10^{-10}$ and $-1.06(14)\times 10^{-10}$
relative to the old KNT19- and DHMZ-based results of Ref.~\cite{spd}.

The impact of the shift in the MI $2\pi$ contribution to
$a_\mu^{\rm HVP}$ from the preliminary result, $3.65(67)\times 10^{-10}$,
employed in Ref.~\cite{spd} to the most recent version, $3.79(19)\times 10^{-10}$,
quoted in Ref.~\cite{Hoferichter:2023sli}, is a very small upward shift of
$0.02\times 10^{-10}$ in the result for $a_\mu^{\rm HVP,s+lqd}$
obtained in Ref.~\cite{spd}. The increase from the initially published
result, $3.68(17)\times 10^{-10}$ \cite{Colangelo:2022prz}, to the updated
$3.79(19)\times 10^{-10}$ version \cite{Hoferichter:2023sli}, similarly,
produces a downward shift of $0.12\times 10^{-10}$ in the result for
$a_\mu^{\rm HVP,lqc}$ obtained in Ref.~\cite{lqc}. The related increase of
the MI $2\pi$ contribution to $a_\mu^{W1}$ from the initially published
$0.83(6)\times 10^{-10}$ result \cite{Colangelo:2022prz} to the recently
updated result, $0.86(6)\times 10^{-10}$ \cite{Hoferichter:2023sli},
similarly produces a downward shift of $0.03\times 10^{-10}$ in the result
for $a_\mu^{\rm W1,lqc}$ obtained in Ref.~\cite{Benton:2023dci}.

We turn, finally, to the impact of the improved determination
of the MI $3\pi$ contribution of Ref.~\cite{Hoferichter:2023bjm} on
the results for $a_\mu^{\rm HVP,s+lqd}$ obtained in Ref.~\cite{spd}.
In Ref.~\cite{spd}, the MI $3\pi$ correction was estimated based on
a VMD model fit by BaBar to BaBar $e^+ e^-\rightarrow 3\pi$ cross
sections \cite{BABAR:2021cde}. The model employed involved an amplitude
consisting of a sum of nominally isospin-conserving (IC) $\omega$, $\phi$
and excited $\omega$ contributions, each proportional to the corresponding
propagator, and an IB $\rho$ contribution proportional to the $\rho$
propagator. The MI $3\pi$ correction to $a_\mu^{\rm HVP,s+lqd}$ was
estimated using results provided by BaBar for the contributions to
$a_\mu^{\rm HVP}$ obtained using the fitted VMD form with and without
the $\rho$ contribution included. The IB $\rho$ contribution in the
VMD model used by BaBar, however, does not have the $\rho$-$\omega$
mixing form, and hence presumably represents the $\rho$ part of the
partial-fraction decomposition of the underlying IB mixing-induced form.
That partial-fraction decomposition would also produce a second IB
contribution, proportional to the $\omega$ propagator, the effect of
which, in the BaBar model, would be absorbed into the fitted strength
of the nominally IC $\omega$ contribution to the amplitude. The squared
modulus of the $\omega$ amplitude contribution to BaBar's fitted VMD
representation of the cross sections will thus contain a hidden IB part
resulting from the interference of this IB contribution with the
corresponding IC part of the $\omega$ contribution to the amplitude.
This hidden contribution is missing from the BaBar-fit-based estimate
of the MI $3\pi$ contribution to $a_\mu^{\rm HVP}$ employed in Ref.~\cite{spd},
but automatically taken into account in the form used in determining that
contribution in Refs.~\cite{Hoferichter:2023bjm,Hoferichter:2023sli}.
We thus replace the BaBar-fit-based estimate with that obtained in
Refs.~\cite{Hoferichter:2023bjm,Hoferichter:2023sli}. This produces
a shift of $+2.12(69)\times 10^{-10}$ in the MI $3\pi$ correction to
$a_\mu^{\rm HVP,s+lqd}$.

Combining the effects of the updates above, we find that
$a_\mu^{\rm HVP,lqc}$ experiences only a very small $0.2\times 10^{-10}$
upward shift, raising the KNT19- and DHMZ-based results of Ref.~\cite{lqc}
from $635.0(2.7)\times 10^{-10}$ to $635.2(2.7)\times 10^{-10}$
and $638.1(4.1)\times 10^{-10}$ to $638.3(4.1)\times 10^{-10}$,
respectively. The effect on $a_\mu^{W1,\rm lqc}$ is even smaller, shifting the
KNT19-based result of Ref.~\cite{Benton:2023dci}, $198.8(1.1)\times 10^{-10}$,
to $198.9(1.1)\times 10^{-10}$. The updates have a somewhat larger effect
on $a_\mu^{\rm HVP,s+lqd}$, with the KNT19- and DHMZ-based results of
Ref.~\cite{spd} shifted upward from $40.1(1.4)(1.3)\times 10^{-10}$ to
$41.1(1.4)(0.4)$ and $38.7(1.4)(1.3)_{\rm lin}(0.4)\times 10^{-10}$
to $39.8(1.4)(1.3)_{\rm lin}(0.4)\times 10^{-10}$, respectively. The
corresponding results for the isospin-limit disconnected contribution to
$a_\mu^{\rm HVP}$, obtained by subtracting the lattice average for the
strange connected contribution from the s+lqd totals, are, using the
same notation for the errors as in Ref.~\cite{spd}, similarly shifted, from
$-13.3(1.4)(0.4)\times 10^{-10}$ to $-12.3(1.4)(0.4)\times 10^{-10}$ and
$-14.6(1.4)(1.3)_{\rm lin}(0.4)\times 10^{-10}$ to
$-13.5(1.4)(1.3)_{\rm lin}(0.4)\times 10^{-10}$.

\begin{table}[t]
\begin{center}
\begin{tabular}{|l|l|l|}
\hline
& $a_\m^{W1,\rm lqc}$ & tension \\
\hline
this work & 198.9(1.1) &  \\
\hline
BMW 20 & 207.3(1.4) & 4.7$\s$\\
LM 20 & 206.0(1.2) & 4.4$\s$ \\
$\c$QCD 23 & 206.7(1.8) & 3.7$\s$ \\
ABGP 22 &206.8(2.2) & 3.2$\s$\\
Mainz/CLS 22 & 207.0(1.5) & 4.4$\s$ \\
ETMC 22 & 206.5(1.3) & 4.5$\s$\\
FHM 23 & 206.6(1.0) & 5.2$\s$\\
RBC/UKQCD 23 & 206.36(0.61) & 5.9$\s$ \\
\hline
\end{tabular}
\begin{quotation}
\floatcaption{comparisontab}%
{{\it Table of the result of Eq.~(\ref{finallqc}) and lattice results
for $a_\m^{W1,\rm lqc}$ from Ref.~\cite{BMW} (BMW 20), Ref.~\cite{Lehner:2020crt}
(LM 20), Ref.~\cite{Wang:2022lkq} ($\c$QCD 23), Ref.~\cite{Aubin:2022hgm} (ABGP 22),
Ref.~\cite{Ce:2022kxy} (Mainz/CLS 22), Ref.~\cite{ExtendedTwistedMass:2022jpw}
(ETMC 22), Ref.~\cite{Bazavov:2023has} (FHM 23), and Ref.~\cite{Blum:2023qou}
(RBC/UKQCD 23). Units of $10^{-10}$. The third column gives the tension
between each of the lattice results and our data-based result, in units
of the error on the difference.}}
\end{quotation}
\end{center}
\end{table}

\vspace{0.8cm}
\section{\label{comp} Comparison with other determinations}
For the lqc $W1$ window quantity, $a_\m^{W1,\rm lqc}$, we compare our
result, Eq.~(\ref{finallqc}), with recent lattice computations in
Table~\ref{comparisontab} and Fig.~\ref{comparison}. We refrain from
quoting a lattice average for $a_\m^{W1,\rm lqc}$,\footnote{We assume
such an average to be forthcoming in an update of the WP, Ref.~\cite{review}.}
but it is clear that there is a discrepancy of about $7\times 10^{-10}$
between the data-based value and lattice results. In the table,
we list the tensions between each of the lattice results, and
the value of Eq.~(\ref{finallqc}). The tensions are significant and range from
$3.2\sigma$ up to $5.9\sigma$.

\begin{table}[t]
\begin{center}
\begin{tabular}{|l|l|}
\hline
this work & 27.0(8) \\
\hline
RBC/UKQCD 18 & 26.0(2) \\
BMW 20 & 26.32(7) \\
Mainz/CLS 22 & 26.87(30) \\
ETMC 22 & 26.50(29) \\
\hline
\end{tabular}
\begin{quotation}
\floatcaption{comparisontabspd}%
{{\it Table of the result of Eq.~(\ref{finalslqd}) and lattice results for
$a_\m^{W1,\rm s+lqd}$ from Ref.~\cite{RBC} (RBC/UKQCD 18), Ref.~\cite{BMW} (BMW 20),
Ref.~\cite{Ce:2022kxy} (Mainz/CLS 22), Ref.~\cite{ExtendedTwistedMass:2022jpw} (ETMC
22). Units of $10^{-10}$.}}
\end{quotation}
\end{center}
\end{table}

We also compare the s+lqd quantity $a_\m^{W1,\rm s+lqd}$ of
Eq.~(\ref{finalslqd}) with results from those collaborations that have
computed $a_\m^{W1,\rm s+lqd}$ on the lattice as well, in
Table~\ref{comparisontabspd} and Fig.~\ref{comparisonslqd}.
The lattice and dispersive results are, in this case, seen to be
compatible within errors, as was the case for the related s+lqd
quantity, $a_\mu^{\rm HVP, s+lqd}$.

Two lattice collaborations have computed $a_\m^{W2,\rm lqc}$, with
Ref.~\cite{Aubin:2022hgm} finding the value $102.1(2.4)\times 10^{-10}$, and
Ref.~\cite{Bazavov:2023has} finding the value $100.7(3.2)\times 10^{-10}$.
This is to be compared with the data-based value
$93.75(36)\times 10^{-10}$ obtained in Eq.~(\ref{finallqcother}), see
Fig.~\ref{comparisonW2}. Our result displays a tension of $3.4\sigma$
with the result of Ref.~\cite{Aubin:2022hgm} and $2.2\sigma$ with the
result of Ref.~\cite{Bazavov:2023has}. 

Finally, up to the pure $I=0$ and $I=1$ EM corrections not included
in Eq.~(\ref{finallqcother}) but expected to be very small, the lqc results of
Eq.~(\ref{finallqcother}) for $I_{\widehat{W}_{15}}^{\rm lqc}$ and
$I_{\widehat{W}_{25}}^{\rm lqc}$ can be compared to the lattice
results obtained from the data of Ref.~\cite{Aubin:2022hgm} in Ref.~\cite{sumrule}:
\begin{eqnarray}
\label{lattIW}
I_{\widehat{W}_{15}}^{\rm lqc}(\mbox{lattice})&=&46.7(0.7)_{\rm stat\ only}
\times 10^{-2}\ ,\\
I_{\widehat{W}_{25}}^{\rm lqc}(\mbox{lattice})&=&82.4(1.0)_{\rm stat\ only}
\times 10^{-3}\ ,
\nonumber
\end{eqnarray}
where the errors are statistical only. These comparison provide
further evidence of tension between lattice and data-driven results
for lqc contributions, though one should keep in mind that the
lattice results were obtained in Ref.~\cite{sumrule} without a detailed
investigation of systematic errors, which was beyond the scope of that paper.

%%%%%%%%%%%%%%%%%%%
\begin{figure}
\vspace*{4ex}
\begin{center}
\includegraphics*[width=12cm]{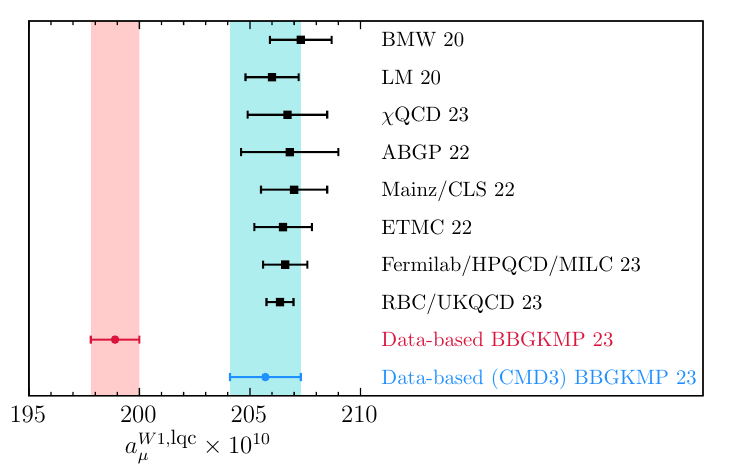}
\end{center}
\begin{quotation}
\floatcaption{comparison}%
{{\it Comparison of the result of Eq.~(\ref{finallqc}) with lattice results
for $a_\m^{W1,\rm lqc}$ from Ref.~\cite{BMW} (BMW 20), Ref.~\cite{Lehner:2020crt}
(LM 20), Ref.~\cite{Wang:2022lkq} ($\c$QCD 23), Ref.~\cite{Aubin:2022hgm} (ABGP 22),
Ref.~\cite{Ce:2022kxy} (Mainz/CLS 22), Ref.~\cite{ExtendedTwistedMass:2022jpw}
(ETMC 22), Ref.~\cite{Bazavov:2023has} (FHM 23), and Ref.~\cite{Blum:2023qou}
(RBC/UKQCD 23). Also shown is the data-based result if the 2-pion data in the
interval between 0.33 and 1.2~GeV is replaced by the results from
CMD-3 \cite{CMD3}.}}
\end{quotation}
\vspace*{-4ex}
\end{figure}
%%%%%%%%%%%%%%%%%%%

%%%%%%%%%%%%%%%%%%%
\begin{figure}
\vspace*{4ex}
\begin{center}
\includegraphics*[width=12cm]{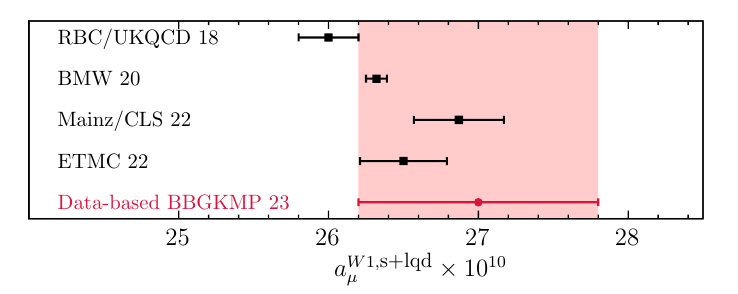}
\end{center}
\begin{quotation}
\floatcaption{comparisonslqd}%
{{\it Comparison of the result of Eq.~(\ref{finalslqd}) with lattice results for
$a_\m^{W1,\rm s+lqd}$ from Ref.~\cite{RBC} (RBC/UKQCD18), Ref.~\cite{BMW} (BMW 20),
Ref.~\cite{Ce:2022kxy} (Mainz/CLS 22) and Ref.~\cite{ExtendedTwistedMass:2022jpw}
(ETMC 22).}}
\end{quotation}
\vspace*{-4ex}
\end{figure}
%%%%%%%%%%%%%%%%%%%

%%%%%%%%%%%%%%%%%%%
\begin{figure}
\vspace*{4ex}
\begin{center}
\includegraphics*[width=12cm]{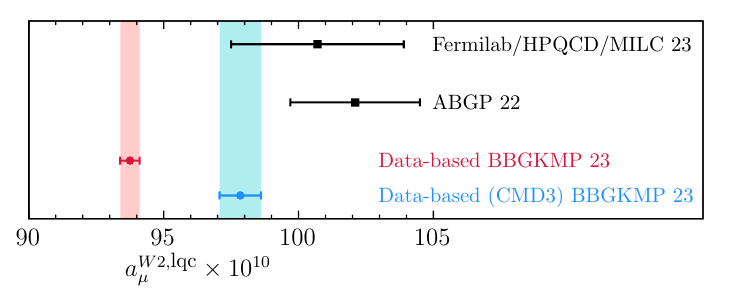}
\end{center}
\begin{quotation}
\floatcaption{comparisonW2}%
{{\it Comparison of the result of Eq.~(\ref{finallqc}) with lattice results
for $a_\m^{W2,\rm lqc}$ from Ref.~\cite{Aubin:2022hgm} (ABGP 22) and
 Ref.~\cite{Bazavov:2023has} (FHM 23). Also shown is the data-based result
if the 2-pion data in the interval between 0.33 and 1.2~GeV is replaced
by the results from CMD-3 \cite{CMD3}.}}
\end{quotation}
\vspace*{-4ex}
\end{figure}
%%%%%%%%%%%%%%%%%%%

\section{\label{concl} Conclusion}
In this paper we have obtained data-driven determinations of
the lqc and s+lqd contributions to a number of window quantities.
Data-driven determinations of such quantities require as input
$s$-dependent exclusive-mode distributions, and the results for
those determinations reported here are based solely on KNT19
results for those distributions. It would be of interest to repeat
the analysis with DHMZ exclusive-mode input, should results for
those distributions eventually become publicly available.

Our result for $a_\mu^{W1, \rm s+lqd}$ is in good agreement with
lattice determinations of this quantity. Similar agreement was
found previously for $a_\mu^{\rm HVP, s+lqd}$ in Ref.~\cite{spd}. These
are, at present, the only quantities for which lattice s+lqd results
exist. It would be of interest to have lattice results, and carry
out analogous comparisons, for the other s+lqd quantities considered
here.

In contrast to the s+lqd case, our results for the lqc contributions
to all four window quantities show tensions with corresponding lattice
results. This tension is particularly significant for $a_\mu^{W1, {\rm lqc}}$,
where, for example, our result differs by $5.9\sigma$ from that of
Ref.~\cite{Blum:2023qou}. Improved lattice determinations of
$a_\mu^{W2, {\rm lqc}}$,
$I_{\widehat{W}_{15}}^{\rm lqc}$ and $I_{\widehat{W}_{25}}^{\rm lqc}$
would be of interest for exploring further the tensions in these
cases, especially so for $I_{\widehat{W}_{15}}^{\rm lqc}$ and
$I_{\widehat{W}_{25}}^{\rm lqc}$ where, at present, only statistical
errors are available for the lattice results.

A final issue of relevance to assessing the significance of the
observed lqc discrepancies is the potential impact of recent
CMD-3 results for the $e^+e^-\rightarrow \pi^+\pi^-$ cross
sections \cite{Blum:2023qou}. As is well known, the results are
significantly higher than those of earlier experiments in the
$\rho$ peak region and, were the CMD-3 results to be correct,
the resulting change in the dispersive evaluation of $a_\mu^{\rm HVP}$
would essentially remove the current discrepancy between the Standard
Model expectation and experimental result for $a_\mu$. The
discrepancies between the new CMD-3 results and those of the
earlier experiments are, however, sufficiently large that a
convincing combination of all existing results does not, at present,
seem possible. Given the unsettled experimental situation, we can
carry out only a preliminary exploration of the potential impact
of the new CMD-3 results. This has been done by replacing the KNT19
$2\pi$ contributions to $R(s)$ in the region covered by CMD-3 data
($E_{\rm CM}$ from $0.327$ to $1.199$ GeV) with the corresponding
contributions implied by CMD-3 data alone. This requires applying
vacuum polarization (VP) corrections to the physical cross sections
implied by the results for the physical timelike pion form factor
quoted by CMD-3 and dressing the resulting bare cross sections with
the final state radiation (FSR) correction factors used by CMD-3
in their evaluation of the contribution of their results to
$a_\mu^{\rm HVP}$. We have used the same VP corrections and same
FSR dressing factors  as those employed by CMD-3.{\footnote{We
thank Fedor Ignatov for providing a link to the file containing
the VP correction results.}} The lqc results produced by this
modification of the $2\pi$ distribution, of course, constitute
only very preliminary explorations, and should in no way be
interpreted as resulting from the use of some updated combination
of the $2\pi$ data base, which no one at present knows how to carry
out. The results of this (we again emphasize {\it preliminary})
exploration are shown for $a_\mu^{W1, {\rm lqc}}$ and $a_\mu^{W2, {\rm lqc}}$
in Figs. \ref{comparison} and \ref{comparisonW2}. As found in the
case of $a_\mu^{\rm HVP}$, use of the CMD-3 $2\pi$ data alone in the
region where it exists removes essentially the entirety of the
observed lqc lattice-data-driven discrepancies.

While the experimental discrepancy between the CMD-3 data and
other data sets for $e^+e^-\to$\ hadrons remains unresolved at present,
we conclude that there are significant discrepancies between the
light-quark-connected parts of all window quantities investigated in this
paper as obtained from the KNT19 compilation of these 
other data sets and recent lattice results, with lattice values pointing to a
value for $a_\mu^{\rm HVP}$ that would bring the SM expectation
for $a_\mu$ much closer to the experimental value.
Further lattice computations of $a_\m^{\rm W2,lqc}$ in particular would 
increase our understanding of the 
discrepancy for this quantity discussed in Sec.~\ref{comp}.\\

%\newpage
\noindent {\bf Acknowledgments}\\

We would like to thank Martin Hoferichter and Peter Stoffer for extensive
discussions on isospin breaking, as well as for providing us with the
$2\p$ mixed-isospin contributions to the $W2$, $\widehat{W}_{15}$ and
$\widehat{W}_{25}$ windows. We also thank Martin Hoferichter and collaborators
for providing the $3\p$ mixed-isospin contributions to those window
quantities, and Martin Hoferichter for useful comments on the $\p^0\g$ and $\h\g$ 
channels. DB and KM thank San Francisco State University where part of
this work was carried out, for hospitality.
This material is based upon work supported by the U.S. Department
of Energy, Office of Science, Office of Basic Energy Sciences Energy
Frontier Research Centers program under Award Number DE-SC-0013682 (GB and MG).
DB's work was supported by the S\~ao Paulo Research Foundation (FAPESP)
Grant No. 2021/06756-6 and by CNPq Grant No. 308979/2021-4.
The work of AK is supported by The Royal Society (URF{$\backslash$}R1{$\backslash$}231503), STFC (Consolidated Grant ST/S000925/) and the European Union's Horizon 2020 research and innovation programme under the Marie Sklodowska-Curie grant agreement No. 858199 (INTENSE).  The work of KM is supported by a grant from the
Natural Sciences and Engineering Research Council of Canada. SP is
supported by the Spanish Ministry of Science, Innovation and Universities
(project PID2020-112965GB-I00/AEI/10.13039/501100011033) and by Departament
de Recerca i Universitats de la Generalitat de Catalunya, Grant No 2021
SGR 00649. IFAE is partially funded by the CERCA program of the Generalitat
de Catalunya.

\appendix
\section{\label{otherweights} Isospin tables for \begin{boldmath} $a_\m^{W2}$, $I_{\widehat{W}_{15}}$ and
$I_{\widehat{W}_{15}}$ \end{boldmath} and further G-parity-ambiguous exclusive-mode contributions}
In this appendix, we show, in Tables~\ref{tab4}, \ref{tab5} and
\ref{tab6}, respectively, the G-parity-unambiguous $I=1$ and $I=0$
exclusive-mode contributions to $a_\m^{W2}$, $I_{\widehat{W}_{15}}$ and
$I_{\widehat{W}_{25}}$. We also list, in Eqs.~(\ref{KKbar3pilqc}) to
~(\ref{gammaspd}), the G-parity-ambiguous exclusive residual-mode
contributions to the lqc and s+lqd parts of $a_\m^{W1}$, $a_\m^{W2}$,
$I_{\widehat{W}_{15}}$ and $I_{\widehat{W}_{25}}$, using the maximally
conservative split prescription of Eq.~(\ref{maxconserv}).

\begin{table}[t]
\begin{center}
\begin{tabular}{lr|lr}
\hline
$I=1$ modes $X$&$[a_\mu^{W2}]_X\times 10^{10}$&$I=0$ modes $X$&
$[a_\mu^{W2}]_X\times 10^{10}$\\
\hline
low-$s$ $\pi^+ \pi^-$& 0.09(02)\quad& low-$s$ $3\pi$& 0.00(00)\quad\\
$\pi^+ \pi^-$& 84.20(33)\quad& $3\pi$& 6.20(14)\quad\\
$2\pi^+ 2\pi^-$& 0.26(00)\quad& $2\pi^+2\pi^-\pi^0$ (no $\omega$,
$\eta$)& 0.01(00)\quad\\
$\pi^+ \pi^- 2\pi^0$& 0.44(02)\quad& $\pi^+ \pi^- 3\pi^0$ (no $\eta$)&
0.00(00)\quad\\
$3\pi^+ 3\pi^-$ (no $\omega$)& 0.00(00)\quad& $3\pi^+3\pi^- \pi^0$ (no
$\omega$, $\eta$)& 0.00(00)\quad\\
$2\pi^+2\pi^-2\pi^0$ (no $\eta$)& 0.00(00)\quad& $\eta \pi^+\pi^-\pi^0$
(no $\omega$)&0.00(00)\quad\\
$\pi^+\pi^- 4\pi^0$ (no $\eta$)& 0.00(00)\quad& $\eta \omega$& 0.00(00)\quad\\
$\eta \pi^+ \pi^-$& 0.01(00)\quad& $\omega (\rightarrow npp )2\pi$&
0.00(00)\quad\\
$\eta 2\pi^+ 2\pi^-$& 0.00(00)\quad& $\omega 2\pi^+ 2\pi^-$&0.00(00)\quad\\
$\eta \pi^+\pi^- 2\pi^0$& 0.00(00)\quad& $\eta \phi$& 0.00(00)\quad\\
$\omega (\rightarrow \pi^0\gamma)\pi^0$& 0.03(00)\quad& $\phi \rightarrow
unaccounted$ &  0.00(00)\quad\\
$\omega (\rightarrow npp)3\pi$& 0.00(00)\quad& \\
$\omega \eta \pi^0$& 0.00(00)\quad&
\\
\hline
TOTAL:& 85.05(33)\quad&TOTAL:& 6.23(14)\quad\\
\hline
\end{tabular}
\floatcaption{tab4}{\it $G$-parity-unambiguous exclusive-mode contributions
to $a_\m^{W2}$ for $\sqrt{s}\leq 1.937$~{\rm GeV} using KNT19 exclusive-mode
data. Entries in units of $10^{-10}$. The notation ``npp'' is KNT shorthand
for ``non-purely-pionic.''}
\end{center}
\end{table}

\begin{table}[ht]
\begin{center}
\begin{tabular}{lr|lr}
\hline
$I=1$ modes $X$&$[I_{\widehat{W}_{15}}]_X\times 10^2$&$I=0$ modes $X$&
$[I_{\widehat{W}_{15}}]_X\times 10^2$\\
\hline
low-$s$ $\pi^+ \pi^-$& 0.00(00)\quad& low-$s$ $3\pi$& 0.00(00)\quad\\
$\pi^+ \pi^-$& 39.35(14)\quad& $3\pi$& 3.83(08)\quad\\
$2\pi^+ 2\pi^-$& -0.02(00)\quad& $2\pi^+2\pi^-\pi^0$ (no $\omega$,
$\eta$)& -0.02(00)\quad\\
$\pi^+ \pi^- 2\pi^0$& 0.10(01)\quad& $\pi^+ \pi^- 3\pi^0$ (no $\eta$)&
-0.01(00)\quad\\
$3\pi^+ 3\pi^-$ (no $\omega$)& -0.01(00)\quad& $3\pi^+3\pi^- \pi^0$ (no
$\omega$, $\eta$)& 0.00(00)\quad\\
$2\pi^+2\pi^-2\pi^0$ (no $\eta$)& -0.04(00)\quad& $\eta \pi^+\pi^-\pi^0$
(no $\omega$)& -0.02(00)\quad\\
$\pi^+\pi^- 4\pi^0$ (no $\eta$)& -0.01(01)\quad& $\eta \omega$& -0.01(00)
\quad\\
$\eta \pi^+ \pi^-$& -0.02(00)\quad& $\omega (\rightarrow npp )2\pi$&
0.00(00)\quad\\
$\eta 2\pi^+ 2\pi^-$& 0.00(00)\quad& $\omega 2\pi^+ 2\pi^-$&
0.00(00)\quad\\
$\eta \pi^+\pi^- 2\pi^0$& 0.00(00)\quad& $\eta \phi$& -0.01(00)\quad\\
$\omega (\rightarrow \pi^0\gamma)\pi^0$& 0.02(00)\quad& $\phi \rightarrow
unaccounted$ &  0.00(00)\quad\\
$\omega (\rightarrow npp)3\pi$& 0.00(00)\quad& \\
$\omega \eta \pi^0$& -0.01(00)\quad& \\
\hline
TOTAL:& 39.37(14)\quad&TOTAL:& 3.77(08)\quad\\
\hline
\end{tabular}
\floatcaption{tab5}{\it $G$-parity-unambiguous exclusive-mode contributions
to $I_{\widehat{W}_{15}}$ for $\sqrt{s}\leq 1.937$~{\rm GeV} using KNT19
exclusive-mode data. Entries in units of $10^{-2}$. The notation ``npp''
is KNT shorthand for ``non-purely-pionic.''}
\end{center}
\end{table}

\begin{table}[ht]
\begin{center}
\begin{tabular}{lr|lr}
\hline
$I=1$ modes $X$&$[I_{\widehat{W}_{25}}]_X\times 10^3$&$I=0$ modes $X$&
$[I_{\widehat{W}_{25}}]_X\times 10^3$\\
\hline
low-$s$ $\pi^+ \pi^-$& 0.00(00)\quad& low-$s$ $3\pi$& 0.00(00)\quad\\
$\pi^+ \pi^-$& 56.60(19)\quad& $3\pi$& 8.05(15)\quad\\
$2\pi^+ 2\pi^-$& 4.13(06)\quad& $2\pi^+2\pi^-\pi^0$ (no $\omega$,
$\eta$)& 0.27(02)\quad\\
$\pi^+ \pi^- 2\pi^0$& 5.31(21)\quad& $\pi^+ \pi^- 3\pi^0$ (no $\eta$)&
0.17(03)\quad\\
$3\pi^+ 3\pi^-$ (no $\omega$)& 0.06(00)\quad& $3\pi^+3\pi^- \pi^0$ (no
$\omega$, $\eta$)& 0.00(00)\quad\\
$2\pi^+2\pi^-2\pi^0$ (no $\eta$)& 0.36(05)\quad& $\eta \pi^+\pi^-\pi^0$
(no $\omega$)&0.19(02)\quad\\
$\pi^+\pi^- 4\pi^0$ (no $\eta$)& 0.06(06)\quad& $\eta \omega$&
0.08(01)\quad\\
$\eta \pi^+ \pi^-$& 0.37(01)\quad& $\omega (\rightarrow npp )2\pi$&
0.04(00)\quad\\
$\eta 2\pi^+ 2\pi^-$& 0.02(00)\quad& $\omega 2\pi^+ 2\pi^-$&
0.00(00)\quad\\
$\eta \pi^+\pi^- 2\pi^0$& 0.03(01)\quad& $\eta \phi$& 0.11(01)\quad\\
$\omega (\rightarrow \pi^0\gamma)\pi^0$& 0.24(01)\quad& $\phi \rightarrow
unaccounted$ &  0.01(01)\quad\\
$\omega (\rightarrow npp)3\pi$& 0.05(01)\quad& \\
$\omega \eta \pi^0$& 0.06(01)\quad& \\
\hline
TOTAL:& 67.29(30)\quad&TOTAL:& 8.92(16)\quad\\
\hline
\end{tabular}
\floatcaption{tab6}{\it $G$-parity-unambiguous exclusive-mode contributions
to $I_{\widehat{W}_{25}}$ for $\sqrt{s}\leq 1.937$~{\rm GeV} using KNT19
exclusive-mode data. Entries in units of $10^{-3}$. The notation ``npp''
is KNT shorthand for ``non-purely-pionic.''}
\end{center}
\end{table}

\begin{itemize}
\item[$\circ$] $K\bar{K}3\p$ modes:
\begin{eqnarray}
\label{KKbar3pilqc}
\left[a_\m^{W1,\rm lqc}\right]_{K\bar{K}3\p}&=&0.012(12)\times 10^{-10}
\ ,\\
\left[a_\m^{W2,\rm lqc}\right]_{K\bar{K}3\p}&=&0.0000(0)\times 10^{-10}
\ ,\nonumber\\
\left[I_{\widehat{W}_{15}}^{\rm lqc}\right]_{K\bar{K}3\p}&=&
-0.00077(77)\times 10^{-2}
\ ,\nonumber\\
\left[I_{\widehat{W}_{25}}^{\rm lqc}\right]_{K\bar{K}3\p}&=&0.0050(50)
\times 10^{-3}
\ .\nonumber
\end{eqnarray}

\begin{eqnarray}
\label{KKbar3pispd}
\left[a_\m^{W1,\rm s+lqd}\right]_{K\bar{K}3\p}&=&0.010(12)\times 10^{-10}
\ ,\\
\left[a_\m^{W2,\rm s+lqd}\right]_{K\bar{K}3\p}&=&0.0000(0)\times 10^{-10}
\ ,\nonumber\\
\left[I_{\widehat{W}_{15}}^{\rm s+lqd}\right]_{K\bar{K}3\p}&=&
-0.00062(77)\times 10^{-2}
\ ,\nonumber\\
\left[I_{\widehat{W}_{25}}^{\rm s+lqd}\right]_{K\bar{K}3\p}&=&0.0040(50)
\times 10^{-3}
\ .\nonumber
\end{eqnarray}

\item[$\circ$] $\o(\rightarrow npp) K\bar{K}$-modes:
\begin{eqnarray}
\label{omKKbarlqc}
\left[a_\m^{W1,\rm lqc}\right]_{\o K\bar{K}}&=&0.0012(12)\times 10^{-10}
\ ,\\
\left[a_\m^{W2,\rm lqc}\right]_{\o K\bar{K}}&=&0.0000(0)\times 10^{-10}
\ ,\nonumber\\
\left[I_{\widehat{W}_{15}}^{\rm lqc}\right]_{\o K\bar{K}}&=&
-0.000081(81)\times 10^{-2}
\ ,\nonumber\\
\left[I_{\widehat{W}_{25}}^{\rm lqc}\right]_{\o K\bar{K}}&=&0.00052(52)
\times 10^{-3}
\ .\nonumber
\end{eqnarray}

\begin{eqnarray}
\label{omKKbarspd}
\left[a_\m^{W1,\rm s+lqd}\right]_{\o K\bar{K}}&=&0.0010(12)\times 10^{-10}
\ ,\\
\left[a_\m^{W2,\rm s+lqd}\right]_{\o K\bar{K}}&=&0.0000(0)\times 10^{-10}
\ ,\nonumber\\
\left[I_{\widehat{W}_{15}}^{\rm s+lqd}\right]_{\o K\bar{K}}&=&
-0.000065(81)\times 10^{-2}
\ ,\nonumber\\
\left[I_{\widehat{W}_{25}}^{\rm s+lqd}\right]_{\o K\bar{K}}&=&0.00042(52)
\times 10^{-3}
\ .\nonumber
\end{eqnarray}

\item[$\circ$] $\eta(\rightarrow npp) K\bar{K}$ modes:
\begin{eqnarray}
\label{etaKKbarlqc}
\left[a_\m^{W1,\rm lqc}\right]_{\eta K\bar{K}}&=&0.0050(50)\times 10^{-10}
\ ,\\
\left[a_\m^{W2,\rm lqc}\right]_{\eta K\bar{K}}&=&0.0000(0)\times 10^{-10}
\ ,\nonumber\\
\left[I_{\widehat{W}_{15}}^{\rm lqc}\right]_{\eta K\bar{K}}&=&
-0.00024(24)\times 10^{-2}
\ ,\nonumber\\
\left[I_{\widehat{W}_{25}}^{\rm lqc}\right]_{\eta K\bar{K}}&=&0.0022(22)
\times 10^{-3}
\ .\nonumber
\end{eqnarray}

\begin{eqnarray}
\label{etaKKbarspd}
\left[a_\m^{W1,\rm s+lqd}\right]_{\eta K\bar{K}}&=&0.0040(50)\times 10^{-10}
\ ,\\
\left[a_\m^{W2,\rm s+lqd}\right]_{\eta K\bar{K}}&=&0.0000(0)\times 10^{-10}
\ ,\nonumber\\
\left[I_{\widehat{W}_{15}}^{\rm s+lqd}\right]_{\eta K\bar{K}}&=&
-0.00019(24)\times 10^{-2}
\ ,\nonumber\\
\left[I_{\widehat{W}_{25}}^{\rm s+lqd}\right]_{\eta K\bar{K}}&=&
0.0017(22)\times 10^{-3}
\ .\nonumber
\end{eqnarray}

\item[$\circ$] $p\bar{p}$ and $n\bar{n}$ modes:
\begin{eqnarray}
\label{ppnnlqc}
\left[a_\m^{W1,\rm lqc}\right]_{p\bar{p}+n\bar{n}}&=&0.020(20)\times 10^{-10}
\ ,\\
\left[a_\m^{W2,\rm lqc}\right]_{p\bar{p}+n\bar{n}}&=&0.0000(0)\times 10^{-10}
\ ,\nonumber\\
\left[I_{\widehat{W}_{15}}^{\rm lqc}\right]_{p\bar{p}+n\bar{n}}&=&
-0.0011(11)\times 10^{-2}
\ ,\nonumber\\
\left[I_{\widehat{W}_{25}}^{\rm lqc}\right]_{p\bar{p}+n\bar{n}}&=&0.0083(83)
\times 10^{-3}
\ .\nonumber
\end{eqnarray}

\begin{eqnarray}
\label{ppnnspd}
\left[a_\m^{W1,\rm s+lqd}\right]_{p\bar{p}+n\bar{n}}&=&0.016(20)\times 10^{-10}
\ ,\\
\left[a_\m^{W2,\rm s+lqd}\right]_{p\bar{p}+n\bar{n}}&=&0.0000(0)\times 10^{-10}
\ ,\nonumber\\
\left[I_{\widehat{W}_{15}}^{\rm s+lqd}\right]_{p\bar{p}+n\bar{n}}&=&
-0.001(11)\times 10^{-2}
\ ,\nonumber\\
\left[I_{\widehat{W}_{25}}^{\rm s+lqd}\right]_{p\bar{p}+n\bar{n}}&=&
0.0066(83)\times 10^{-3}
\ .\nonumber
\end{eqnarray}

\item[$\circ$] Low-$s$ $\p^0 \g$ and $\eta \g$ modes:

\begin{eqnarray}
\label{gammalqc}
\left[a_\m^{W1,\rm lqc}\right]_{{\rm low-}s\ \p^0 \g+\eta \g}&=&0.0082(82)
\times 10^{-10}
\ ,\\
\left[a_\m^{W2,\rm lqc}\right]_{{\rm low-}s\ \p^0 \g+\eta \g}&=&0.012(12)
\times 10^{-10}
\ ,\nonumber\\
\left[I_{\widehat{W}_{15}}^{\rm lqc}\right]_{{\rm low-}s\ \p^0 \g+\eta \g}&=&
0.0030(30)\times 10^{-2}
\ ,\nonumber\\
\left[I_{\widehat{W}_{25}}^{\rm lqc}\right]_{{\rm low-}s\ \p^0 \g+\eta \g}&=&
0.0020(20)\times 10^{-3}
\ .\nonumber
\end{eqnarray}

\begin{eqnarray}
\label{gammaspd}
\left[a_\m^{W1,\rm s+lqd}\right]_{{\rm low-}s\ \p^0 \g+\eta \g}&=&
0.0066(82)\times 10^{-10}
\ ,\\
\left[a_\m^{W2,\rm s+lqd}\right]_{{\rm low-}s\ \p^0 \g+\eta \g}&=&
0.010(12)\times 10^{-10}
\ ,\nonumber\\
\left[I_{\widehat{W}_{15}}^{\rm s+lqd}\right]_{{\rm low-}s\ \p^0 \g+\eta\g}&=&
 0.0024(30)\times 10^{-2}
\ ,\nonumber\\
\left[I_{\widehat{W}_{25}}^{\rm s+lqd}\right]_{{\rm low-}s\ \p^0 \g+\eta \g}&=&
0.0016(20)\times 10^{-3}
\ .\nonumber
\end{eqnarray}

\end{itemize}

\section{\label{PgammaVMDdecomp} The VMD representations of the \begin{boldmath} $e^+e^-\rightarrow\pi^0\gamma$
and $e^+e^-\rightarrow\eta\gamma$ \end{boldmath} cross sections}
The cross sections for $e^+ e^-\rightarrow P\gamma$, $P=\pi^0 ,\, \eta$,
$\sigma_{P\gamma}(s)$, are given by
\begin{equation}
\sigma_{P\gamma}(s)={\frac{2\pi^2\alpha_{\rm EM}^3}{3}}
\, \left(1-{\frac{m_P^2}{s}}\right)^3
\, \vert F_{P\gamma^*\gamma^*}(s,0)\vert^2\, ,
\label{basicxsec}\end{equation}
where $F_{P\gamma^*\gamma^*}(s,0)$ is the timelike $P=\pi^0$ or $\eta$
transition form factor. Here, and in what follows, we employ the notation of
Sec.~6 of the recent review, Ref.~\cite{Gan:2020aco}. In this notation, the VMD
representation of the experimentally dominant isoscalar contribution to
$F_{P\gamma^*\gamma^*}(s,0)$ is
\begin{equation}
F^{I=0}_{P\gamma\gamma}\sum_{V=\omega ,\phi} {\frac{w_{PV\gamma}s}
{ m_V^2-s-im_V\Gamma_V(s)}}\, ,
\label{ieq0transitionff}\end{equation}
where $F_{P\gamma\gamma}$ is related to $\Gamma [P\rightarrow\gamma\gamma ]$
by
\begin{equation}
\Gamma [P\rightarrow\gamma\gamma ]={\frac{\pi\alpha_{\rm EM}^2 m_P^3}{4}}
\vert F_{P\gamma\gamma}\vert^2\ ,
\label{twophotongamma}\end{equation}
and the weights $w_{PV\gamma}$ accompanying the $\omega$ and $\phi$
propagators are given by
\begin{equation}
w^2_{PV\gamma} = {\frac{9 m_V^2 m_P^3\Gamma [V\rightarrow e^+ e^-]
\Gamma [V\rightarrow P\gamma ]}{2\alpha_{\rm EM}\left(m_V^2-m_P^2\right)^3
\gamma [P\rightarrow\gamma\gamma ]}}\ .
\label{w2PVgammaform}\end{equation}
In what follows we use $s$-dependent versions of the widths $\Gamma_V(s)$,
following the treatment of Refs.~\cite{SND:2016drm,Achasov:2003ir,Achasov:1989mh},
which takes into account all decay modes with branching fractions greater
than $1\%$. A similar VMD form,
\begin{equation}
F^{I=1}_{P\gamma\gamma} {\frac{w_{P\rho\gamma}s}
{ m_\rho^2-s-im_\rho\Gamma_\rho (s)}}
\label{ieq1transitionff}\end{equation}
again with $s$-dependent width, is used to approximate the smaller
isovector contribution to the transition form factor, with the weights
$w_{P\rho\gamma}$ also given by the general expression in
Eq.~(\ref{w2PVgammaform}). The full VMD representation of the cross
section then becomes
\begin{equation}
\sigma_{P\gamma}(s)={\frac{2\pi^2\alpha_{\rm EM}^3}{3}}
\, \left(1-{\frac{m_P^2}{s}}\right)^3 \vert F_{P\gamma\gamma}\vert^2
\, \vert \sum_{V=\rho ,\omega ,\phi} w_{PV\gamma}\, s/
[m_V^2-s-im_V\Gamma_V(s)]\vert^2\, .
\label{basicxsecfull}\end{equation}
An alternate form of this expression, obtained by pulling out the
$V$-independent factor $m_P/\left[ 2\alpha_{\rm EM} \Gamma
[P\rightarrow\gamma\gamma]\right]$
from the squared modulus in Eq.~(\ref{basicxsecfull}) and substituting the RHS
of Eq.~(\ref{twophotongamma}) for $\Gamma [P\rightarrow\gamma\gamma]$, is
\begin{equation}
\sigma_{P\gamma}(s)={\frac{4\pi}{3 m_P^2}}
\, \left(1-{\frac{m_P^2}{s}}\right)^3 
\, \vert \sum_{V=\rho ,\omega ,\phi} \hat{w}_{PV\gamma}\, s/
[m_V^2-s-im_V\Gamma_V(s)]\vert^2\, ,
\label{altbasicxsec}\end{equation}
where the alternate weights, $\hat{w}_{PV\gamma}$, are given by
\begin{equation}
\hat{w}^2_{PV \gamma} = 9 m_V^2 m_P^2 \Gamma [V\rightarrow e^+ e^-]
\Gamma [V\rightarrow P \gamma ] /\left( m_V^2-m_P^2\right)^3\, .
\label{what2PVgamma}\end{equation}
This alternate form makes more explicit the fact that the contributions
to the amplitude are determined entirely by the vector meson masses and
widths, and the strengths of the corresponding $V\rightarrow e^+ e^-$
and $V\rightarrow P\gamma$ couplings.

Using PDG \cite{Workman:2022ynf} input, we find the following
values for the weights $\hat{w}_{PV\gamma}$,
\begin{eqnarray}
&&\hat{w}_{\pi^0\rho\gamma}=1.559(133)\times 10^{-5}\ ,\\
&&\hat{w}_{\pi^0\omega\gamma}=1.491(36)\times 10^{-5}\ ,\nonumber\\
&&\hat{w}_{\pi^0\phi\gamma}=-1.068(26)\times 10^{-6}\ ,\nonumber\\
&&\hat{w}_{\eta\rho\gamma}=1.362(48)\times 10^{-4}\ ,\nonumber\\
&&\hat{w}_{\eta\omega\gamma}=1.167(56)\times 10^{-4}\ ,\nonumber\\
&&\hat{w}_{\eta\phi\gamma}=-2.210(41)\times 10^{-5}\ . \nonumber
\label{whatPVgammaPDGvals}\end{eqnarray}
As in Ref.~\cite{Gan:2020aco}, the sign of $w_{\eta\phi\gamma}$ is
chosen negative, based on the discussions of Ref.~\cite{Hanhart:2013vba}
and underlying arguments of the earlier review Ref.~\cite{Landsberg:1985gaz}.
While those arguments do not resolve the choice of sign for
$w_{\pi^0\phi\gamma}$, the results of the SND fit of Ref.~\cite{SND:2016drm}
to the $e^+ e^-\rightarrow \pi^0\gamma$ cross sections clearly favor a
relative negative sign between the $\phi$ and $\rho$/$\omega$ amplitude
contributions. The choice of sign for $\hat{w}_{\pi^0\phi\gamma}$, in any
case, turns out to have only a very small effect on the $I=0$/$I=1$/MI
decompositions of the $\pi^0\gamma$ exclusive-mode contributions to the
integrals we are interested in.{\footnote{Explicitly, the $I=1$
contributions are unaffected, while the $I=0$ and MI contributions to
$a_\mu^{\rm HVP}$ shift from $79.5\%$ and $16.6\%$ of the total when
$w_{\pi^0\phi\gamma}$ is chosen $<0$ to $79.7\%$ and $16.3\%$ when it is
chosen $>0$. Those same contributions, similarly, shift from $79.1\%$ and
$16.7\%$ to $79.3\%$ and
$16.5\%$ for $a_\mu^{W1}$, from $79.9\%$ and $16.2\%$ to $80.1\%$ and
$16.0\%$ for $a_\mu^{W2}$, from $79.8\%$ and $16.3\%$ to $80.0\%$ and
$16.1\%$ for $I_{\widehat{W}_{15}}$ and from $79.0\%$ and $16.8\%$ to
$79.2\%$ an $16.6\%$ for $I_{\widehat{W}_{25}}$.}} 

The above input produces the results quoted in the main text for the
$I=0$, $I=1$ and MI contributions to the exclusive-mode $\pi^0\gamma$
and $\eta\gamma$ weighted integrals of interest in this paper. The
associated errors are those resulting from summing in quadrature the
uncertainties produced by those on the three input quantities,
$\hat{w}_{PV\gamma}$, $V=\rho ,\, \omega ,\, \phi$, for the $P=\pi^0$
and $\eta$ cases, respectively. The resulting VMD-model totals from the
region between $E_{\rm CM}=0.600$ GeV (the lowest $E_{\rm CM}$ in the KNT19
$\pi^0\gamma$ mode data compilation) and a point safely above the last
of the enhanced VMD resonance contributions (which we take to be
$E_{\rm CM}=m_\phi +5\Gamma_\phi$), are $4.44(21)\times 10^{-10}$ for the HVP
case, $1.58(8)\times 10^{-10}$ for the W1 case, $0.68(3)\times 10^{-10}$ for
the W2 case, $0.40(2)\times 10^{-2}$ for the $\widehat{W}_{15}$ case and
$0.68(3)\times 10^{-3}$ for the $\widehat{W}_{25}$. Comparing these
to the corresponding KNT19 exclusive-mode contributions, which are
$4.46(10)\times 10^{-10}$, $1.58(4)\times 10^{-10}$,
$0.69(2)\times 10^{-10}$, $0.40(1)\times 10^{-2}$ and $0.66(2)\times 10^{-3}$,
respectively, we see that the VMD contributions, in all cases, saturate the
KNT19 $\pi^0\gamma$ mode results. A similar saturation of KNT19 results by
the corresponding VMD model results is observed for the $\eta\gamma$
mode as well.

%\newpage
\vspace{3ex}
%%%%%%%%%%%%%%%%%%%%%%%%%%%

\end{document}